\documentclass[useAMS,usenatbib]{mn2e}
\usepackage[english]{babel}

\usepackage{fancyhdr}
\usepackage{amsfonts}
\usepackage{amsmath}
\usepackage{amssymb}
\usepackage{multicol}
\usepackage{layout}
\usepackage{psfig}
\usepackage{graphicx}
\usepackage{subfigure}

\usepackage{times}
\usepackage{natbib}

\newif\ifAMStwofonts
\AMStwofontstrue

\renewcommand{\vec}[1]{\bmath{#1}}
\newcommand{\be}{\begin{equation}}
\newcommand{\ee}{\end{equation}}
\newcommand{\ba}{\begin{eqnarray}}
\newcommand{\ea}{\end{eqnarray}}
\newcommand{\brr}{\begin{array}}
\newcommand{\err}{\end{array}}
\newcommand{\bc}{\begin{center}}
\newcommand{\ec}{\end{center}}

\newcommand{\mincir}{\raise
  -2.truept\hbox{\rlap{\hbox{$\sim$}}\raise5.truept \hbox{$<$}\ }}
\newcommand{\magcir}{\raise
  -2.truept\hbox{\rlap{\hbox{$\sim$}}\raise5.truept \hbox{$>$}\ }}
\newcommand{\siml}{\raise
  -2.truept\hbox{\rlap{\hbox{$\sim$}}\raise5.truept \hbox{$<$}\ }}
\newcommand{\simg}{\raise
  -2.truept\hbox{\rlap{\hbox{$\sim$}}\raise5.truept \hbox{$>$}\ }}

\newcommand{\aj}{AJ}
\newcommand{\apj}{ApJ}
\newcommand{\apjl}{ApJ}

\newcommand{\aap}{A\&A}

\newcommand{\mnras}{MNRAS}

\newcommand{\physrep}{Phys. Rep.}
\newcommand {\apgt} {\ {\raise-.5ex\hbox{$\buildrel>\over\sim$}}\ }
\newcommand {\aplt} {\ {\raise-.5ex\hbox{$\buildrel<\over\sim$}}\ }

\title[Inner density profile of DM haloes] {Constraints on the inner density profile of dark-matter haloes from weak gravitational lensing}  \author[Viola et al.] {M. Viola$^{1}$,
  M. Maturi$^{1}$, M. Bartelmann$^{1}$ \\~\\
  $^1$ Zentrum f\"ur Astronomie,ITA, Universit\"at Heidelberg, Albert-Ueberle-Str.2, 69120 Heidelberg, Germany \\~\\
  (mviola,maturi,mbartelmann@ita.uni-heidelberg.de)\\ \\ }

\date{Accepted 2009 December 3. Received 2009 November 18; in original form 2009 September 24}

\begin{document}
\label{firstpage}
\maketitle

\begin{abstract}

We construct two linear filtering techniques based on weak gravitational lensing to constrain the inner slope $\alpha$ of the density profile of dark-matter haloes. Both methods combine all available information into an estimate of this single number. Under idealised assumptions, $\alpha$ is constrained to $\sim15\%$ if the halo concentration $c$ is known, and to $\aplt 30\%$ if not. We argue that the inevitable degeneracy between density-profile slope and halo concentration cannot be lifted under realistic conditions, and show by means of Fisher-matrix methods which linear combination of $\alpha$ and $c$ is best constrained by our filtering of lensing data. This defines a new parameter, called $P_1$, which is then constrained to $\sim15\%$ for a single massive halo. If the signals of many haloes can be stacked, their density profiles should thus be well constrained by the linear filters proposed here with the advantage, in contrast with strong lensing analysis, to be insensitive to the cluster substructures.

\end{abstract}


\section{Introduction}

Numerical simulations of non-linear structure formation in a broad class of cosmological models, even with different types of power spectra for the dark-matter (DM hereafter) density fluctuations, reveal a typical shape for the density profile of DM haloes. As far as the numerical resolution allows this statement, the density profile begins with at least a mild singularity in the core, then falls off with a relatively flat slope out to a characteristic radius where it gently steepens towards an asymptotic behaviour $\propto r^{-3}$ far away from the core. Do real haloes behave in the same way as theory predicts?

Gravitational lensing should in principle be able to give the cleanest answer to this question. Density profiles in galaxy-sized objects are expected to be modified on small scales by baryonic physics, where they are likely to approach the isothermal density slope $\propto r^{-2}$ instead of the generic DM behaviour. On the mass scale of galaxy groups or clusters, however, baryonic physics should be constrained to the innermost region, leaving the DM density profile almost intact. Galaxy-galaxy lensing seems to show tentative evidence for this expectation \citep{Mandelbaum06} : while the shear profile around low-mass haloes is consistent with an isothermal density profile, it seems to flatten towards the theoretical expectation for DM haloes around high-mass haloes.

The question is important because it aims at a central prediction of non-linear cosmological structure formation. Answering it is complicated by the angular resolution limit of $\aplt 20''$ of weak gravitational lensing, set by the number density of background galaxies, and by the high non-linearity of strong gravitational lensing. In fact, claims that strong gravitational lensing, when combined with stellar dynamics, requires flat halo cores have been made \citep{Sand04} and doubted. In particular \citep{Meneghetti07} showed how the measurement of the inner slope can be systematically underestimated if halo's substructure are not taken into account. A weak lensing analysis, even if observationally more challenging, has the advantage to be almost insensitive to cluster's substructures because of the instrinsic nature of the signal.

Previous studies based on weak lensing have followed an approach where a shear profile was first measured and then fit to the shear profile expected from certain three-dimensional density profiles, thus indirectly constraining the density-profile models. Given the sparseness of lensing information near the core of galaxy groups and clusters, we develop a different approach here. Instead of constraining the shear profile, we only wish to derive a single number from the shear data, namely the slope $\alpha$ of the density profile within the characteristic radius, assuming that the asymptotic outer slope is $-3$.

We pursue this approach with two linear filtering techniques. One of them is specifically constructed below to return $\alpha$ as its only result. It is thus made to combine all available information into its estimate and should thus optimise the significance of the measurement. The other varies the inner slope of the density profile until it finds the maximum signal-to-noise ratio in a given sample of haloes.

We proceed as follows. In Sect.~2, we introduce the weak-lensing properties of the generalised NFW density profile. We develop our linear filters in Sect.~3. There, we also discuss the degeneracy between the central density slope and the concentration parameter of the halo and use the Fisher matrix to find a linear parameter combination which is best constrained by shear measurements. The sensitivity of our method and its limitations are shown and discussed in Sects.~4 and 5, respectively. Section~6 presents our conclusions.
\section{Generalized NFW profile}\label{sec:genNFW}

\begin{figure}
	\centerline{
      \psfig{figure=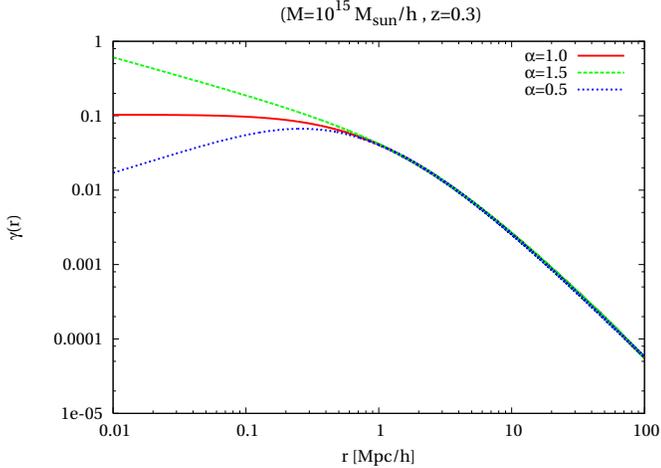,width=9cm,angle=270}}
\caption{Shear profile of a generalized NFW halo for three different values of $\alpha$. The case $\alpha=1$ corresponds to the usual NFW profile. Note the strong dependence on the inner slope.}\label{fig:genShear}
\end{figure}

In the last decade a large effort has been devoted to predict the density profile of DM haloes in $\Lambda CDM$ cosmologies. Thanks to increasing resolution of numerical simulations there is now a common agreement \citep{NFW,Moore98}, that DM density profiles can be accurately described by the generalized NFW profile
\begin{equation}
\rho(r)=\frac{\rho_s}{(r/r_{s})^{\alpha}(1+r/r_s)^{3-\alpha}} \, \label{eq:NFWgen}
\end{equation}
with the scale radius $r_{s}$, the inner slope $\alpha$ and the scale density
\begin{equation}
\rho_s=\rho_{crit}(z)\frac{200(3-\alpha)(r_{200}/r_{s})}{3\mathit{_1F_2}(3-\alpha,3-\alpha,4-\alpha,-r_{200}/r_{s})} \ ,
\end{equation}
where $_1F_2(a,b,c,z)$ is a hypergeometric function and $r_{200}$ is the radius enclosing 200 times the critical density of the universe $\rho_{crit}$. The halo concentration is defined as
\begin{equation}
c_{200}=\frac{r_{200}}{r_s}.\label{eq:con}
\end{equation}
Following \cite{Keeton01} we interpret the scale radius as the radius where the density profile reaches slope $-2$, i.e.~$d\ln \rho/d\ln r =-2$. For the profile of equation \ref{eq:NFWgen}
\begin{equation}
r_{-2}=r_s(2-\alpha) \,\label{eq:scaleRadiusGen}
\end{equation}
and thus
\begin{equation}
c_{-2}=\frac{r_{200}}{r_{-2}}=\frac{1}{2-\alpha}c_{200}.\label{eq:conGen}
\end{equation}
For $\alpha=1$ this formula reduces to the standard NFW case \citep{NFW}. The profile is fully characterized when the mass, the redshift, the concentration and the inner slope of the halo are specified. However not all of these parameters are independent. Numerical simulations show that it is possible to define fitting formulae relating the concentration with the mass and the redshift of the halo. In the following we will use the prescription proposed by \cite{Eke01} for this purpose. They also found that for a fixed value of mass and redshift, the concentration approximately follows a log-normal distribution
\begin{equation}
p(c)dc=\frac{1}{\sqrt{2\pi}\sigma_{c}c}\exp\Bigg[-\frac{(\ln c -\ln \bar{c})^2}{2\sigma^2_{c}}\Bigg]d\ln c \;
\end{equation}
where $\sigma_{c}$ is the 1-$\sigma$ deviation of $\Delta(\ln c) \simeq 0.2$ \citep{NFW,Bullock01a}.

\subsection{Weak lensing properties}\label{sec:weaklensing}

This Section summarises the basic weak-lensing concepts that will be used later. For a complete overview we refer to \cite{Bartelmann01}.  An isolated lens with surface mass density $\Sigma(\vec{\theta})$ has the lensing potential
\begin{equation}
\Psi(\vec{\theta})=\frac{4G}{c^2}\frac{D_{l}D_{s}}{D_{ls}}\int d^2\theta^{'}\Sigma(\vec{\theta}^{'})\ln |\vec{\theta}-\vec{\theta}^{'}|, \label{eq:potential}
\end{equation}
where $G$ is the gravitational constant, $c$ is the speed of light, and $D_{l, s, ls}$ are the angular diameter distances between the observer and the lens, the observer and the source and the lens and the source respectively.

Due to the presence of the lens a light ray is deflected by the angle
\begin{equation}
\vec{\zeta}(\vec{\theta})=\nabla \Psi(\vec{\theta}).\label{eq:deflection}
\end{equation}
A source located at the angular position $\vec\beta$ in the sky is seen by the observer at an angular position $\vec\theta$ which are related by the lens equation
\begin{equation}
\vec{\beta}=\vec{\theta}-\vec{\zeta}(\vec{\theta}).\label{eq:lenseq}
\end{equation}
If the source's angular extent is much smaller than the angular scale on which the lens properties change, the lens mapping can be locally linearised and the image distortion is given by the following Jacobian matrix
\begin{equation}
A\equiv\frac{\partial\vec{\beta}}{\partial \vec{\theta}}=\Bigg(\delta_{ij}-\frac{\partial ^2\Psi(\vec{\theta})}{\partial \theta_{i}\partial \theta_{j}}\Bigg)=\left(\begin{array}{cc}
1-\kappa -\gamma_1 & -\gamma_2 \\
-\gamma_2 & 1-\kappa + \gamma_1 \end{array} \right), \label{eq:distortion}
\end{equation}
where
\begin{equation}
\kappa(\vec{\theta})=\frac{\Sigma(\vec{\theta})}{\Sigma_{cr}}=\frac{1}{2}(\Psi_{11}+\Psi_{22})\label{eq:kappa}
\end{equation}
is the \textit{convergence}, i.e.~the surface mass density scaled by the \textit{critical surface mass density}
\begin{equation}
\Sigma_{cr}=\frac{c^2}{4\pi G}\frac{D_{s}}{D_{l}D_{ls}},\label{eq:sigmacr}
\end{equation}
and
\begin{equation}
\gamma_1=\frac{1}{2}(\Psi_{11}-\Psi_{22}),\gamma_2 = \Psi_{12}.
\end{equation}\label{eq:shearcomp}
are the two components of the complex shear $\gamma=\gamma_1+i\gamma_2$. The net result is a distortion and a magnification of the background sources due to the lens gravitational field.

For an axially-symmetric lens, outside critical curves, the distortion is tangential to the line connecting the source and the lens so that the tangential shear is given by
\begin{equation}
\gamma_T=-(\gamma_{1}\cos(2\theta)+\gamma_{2}\sin(2\theta))=-\Re(\gamma e^{2i\theta}) \,
\end{equation}\label{eq:tanshear}
and the shear modulus can be derived from the convergence,
\begin{equation}
|\gamma|(\theta)=\bar{\kappa}(\theta) -\kappa(\theta) \label{eq:gamma_sym} \,
\end{equation}
where $\bar{\kappa}(\theta)$ is the mean surface mass density inside a circle of radius $\theta$ centered on the lens and $\kappa(\theta)$ is the convergence at radius $\theta$.  If $\alpha \neq 1$ it is not possible to find an analytic expression for the shear profile and therefore equation \ref{eq:gamma_sym} has to be computed numerically. We show in figure \ref{fig:genShear} the shear profile for three values of $\alpha$ (0.5,1.0,1.5). The inner shear profile depends sensitively on the inner slope $\alpha$. In the case $\alpha = 1$  the shear is logarithmically divergent for small values of $\theta$ \citep{Bartelmann96}. The divergence is more pronounced for steeper profiles, while the shear profile decreases for $\alpha < 1$ and tends to converge to a finite value even if it is undefined for $\theta=0$.

The shear profile depends on the two parameters of the density profile, of which the concentration depends mildly on the halo redshift. An additional and stronger dependence on halo and source redshifts is introduced through the geometry of the lens system.

The observable lensing signal is the ellipticity of the background galaxies,
\begin{equation}
e^{obs}=\frac{e^{int}+g}{1+g^{*}e^{int}} \,
\end{equation}
where
\begin{equation}
e^{int}\simeq \frac{1-b/a}{1+b/a}\exp(2i\phi) \,
\end{equation}
is their intrinsic ellipticity, $a$ and $b$ are the major and minor axes, respectively, $\phi$ is the orientation angle, $g$ is the \textit{reduced shear} $g=|\gamma|/(1-\kappa)$ and $g^{*}$ is its complex conjugate. In the weak-lensing regime ($\gamma \ll 1$), $e^{obs} \simeq g \simeq \gamma$. In the following, we will exclusively use the reduced shear since we want to explore scales where the approximation $\kappa \ll 1$ does not hold. Nonetheless, we shall denote it by $\gamma$ throughout for simplicity of notation.

\section{Methods to characterize the shear profile}

In this section we describe two methods, based on optimal linear filters \citep{Sanz01,Maturi05}, to estimate the inner slope of DM haloes using weak-lensing observations. The advantage of linear filtering as opposed to standard profile fitting is that filters can be constructed such as to minimise noise caused by intervening structures along the line-of-sight.

\subsection{Optimal linear filtering}\label{sec:LF}

For a generic optimal linear filter, the data $D(\vec{\theta})$ is modelled as the sum of the signal to be measured and the noise
\begin{equation}
D(\vec{\theta})=S(\vec{\theta})+N(\vec{\theta})\label{eq:data},
\end{equation}
where $S(\vec{\theta})=A\tau(\vec{\theta})$, $A$ is the signal amplitude and $\tau(\vec{\theta})$ is a model for its angular shape. In our application, the signal is the lensing shear of the intervening DM halo and the noise is given by the intrinsic ellipticity of the background galaxies, their finite number and the contamination due to
large-scale structures. The noise components are assumed to be Gaussian, random with zero mean and isotropic since their statistical properties are independent of the position in the sky (for further detail see \citep{Maturi05}).  We now define a linear filter $\Psi(\vec{\theta},\alpha, \vec{w})$ which, when convolved with the data, yields an estimate for the amplitude of the signal at the position $\vec{\theta}$:
\begin{equation}
A_{est}(\vec{\theta})=\int D(\vec{\theta}^{\prime})\Psi(\vec{\theta}-\vec{\theta}^{\prime},\alpha,\vec{w})d^2\theta^{\prime}, \label{aest_fltr}
\end{equation}
which is unbiased
\begin{equation}
b=A\Bigg[\int \Psi(\vec{\theta},\alpha, \vec{w})\tau(\vec{\theta}, \alpha, \vec{w})d^2\theta -1 \Bigg]=0 \,\label{bias_fltr}
\end{equation}
and whose variance $\sigma^2$
\begin{equation}
\sigma^2=b^2+\frac{1}{2\pi}\Bigg[\int |\hat{\Psi}(\vec{k},\alpha,\vec{w})|^2P_N(k)d^2k\Bigg], \label{var_flt}
\end{equation}
is minimal. The filter $\Psi$ satisfying these two conditions minimises the Lagrangian $L=\sigma^2+\lambda b$. It reads
\begin{equation}
\hat{\Psi}(\vec{k})=\frac{1}{2\pi}\Bigg[\int\frac{|\hat{\tau}(\vec{k},\alpha,\vec{w})|^2}{P_{N}(k)}d^{2}k\Bigg]^2 \frac{\hat{\tau}(\vec{k},\alpha,\vec{w})}{P_{N}(k)}.\label{eq:fltr}
\end{equation}
where $\vec{w}=(c,M,z)$ and $\hat{\Psi}$ and $\hat{\tau}$ are the Fourier transforms of the filter and the signal shape, respectively . Note that we have assumed in the previous derivation that the mean values of the halo parameters ($\vec{w}$) are well known. This is an idealising assumption and we refer to Sect.~\ref{sec:mod_sens} for a more detailed discussion.
The filter depends only on the angular shape of the signal $\tau(\vec{k},\alpha, \vec{w})$ and the noise power spectrum $P_N$. In particular it is most sensitive to those spatial frequencies for which the signal $\tau$ is large and the noise power spectrum is small. This filter is optimal in the sense that it maximises the signal-to-noise ratio for the a given assumed signal shape.

\begin{figure*}
  \begin{center}
  \begin{minipage}{180mm}
    \begin{tabular}{cc}
    \resizebox{90mm}{!}{\includegraphics[width=7.0cm,angle=270]{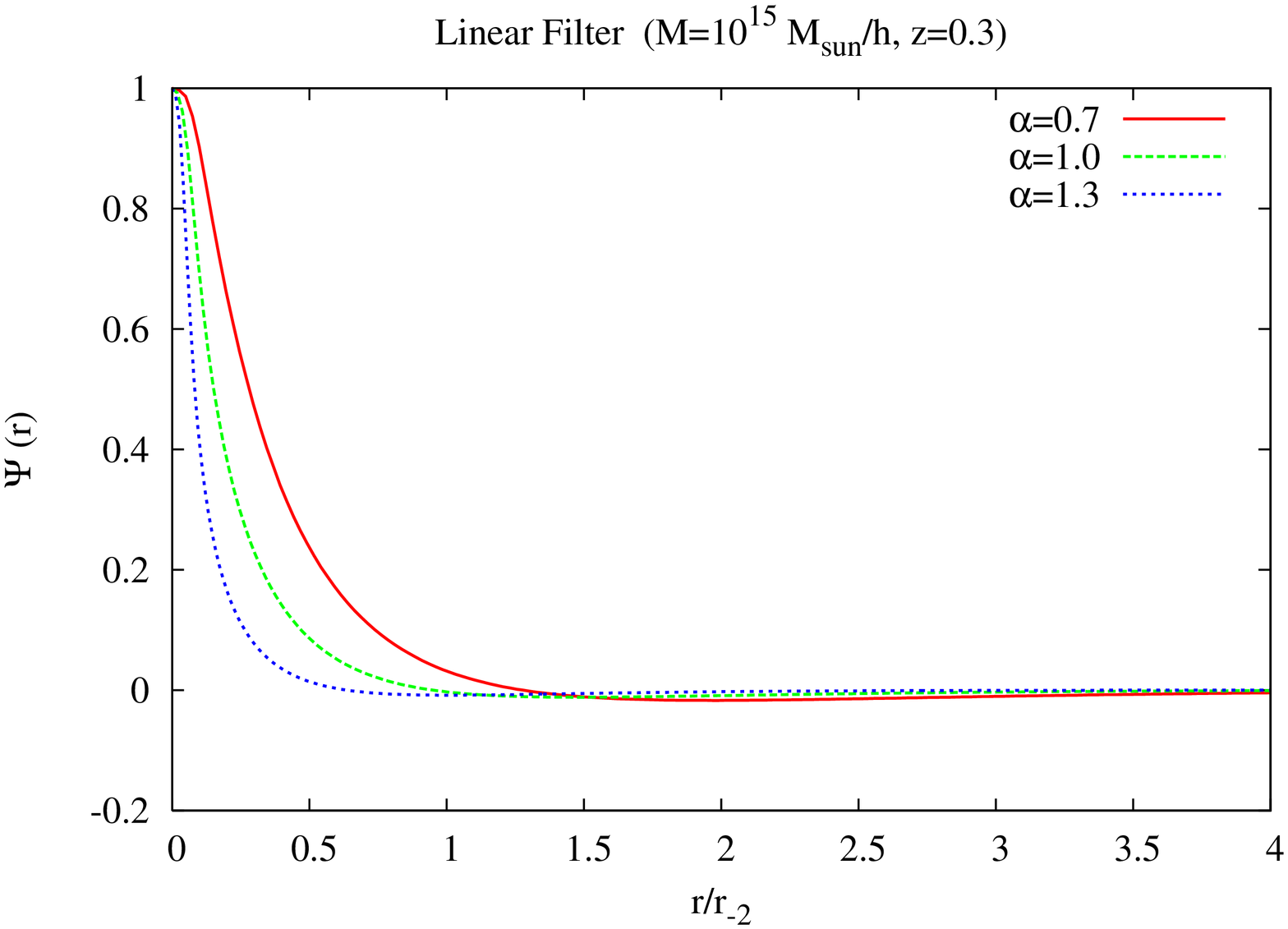}} 
   \resizebox{90mm}{!}{\includegraphics[width=7.0cm,angle=270]{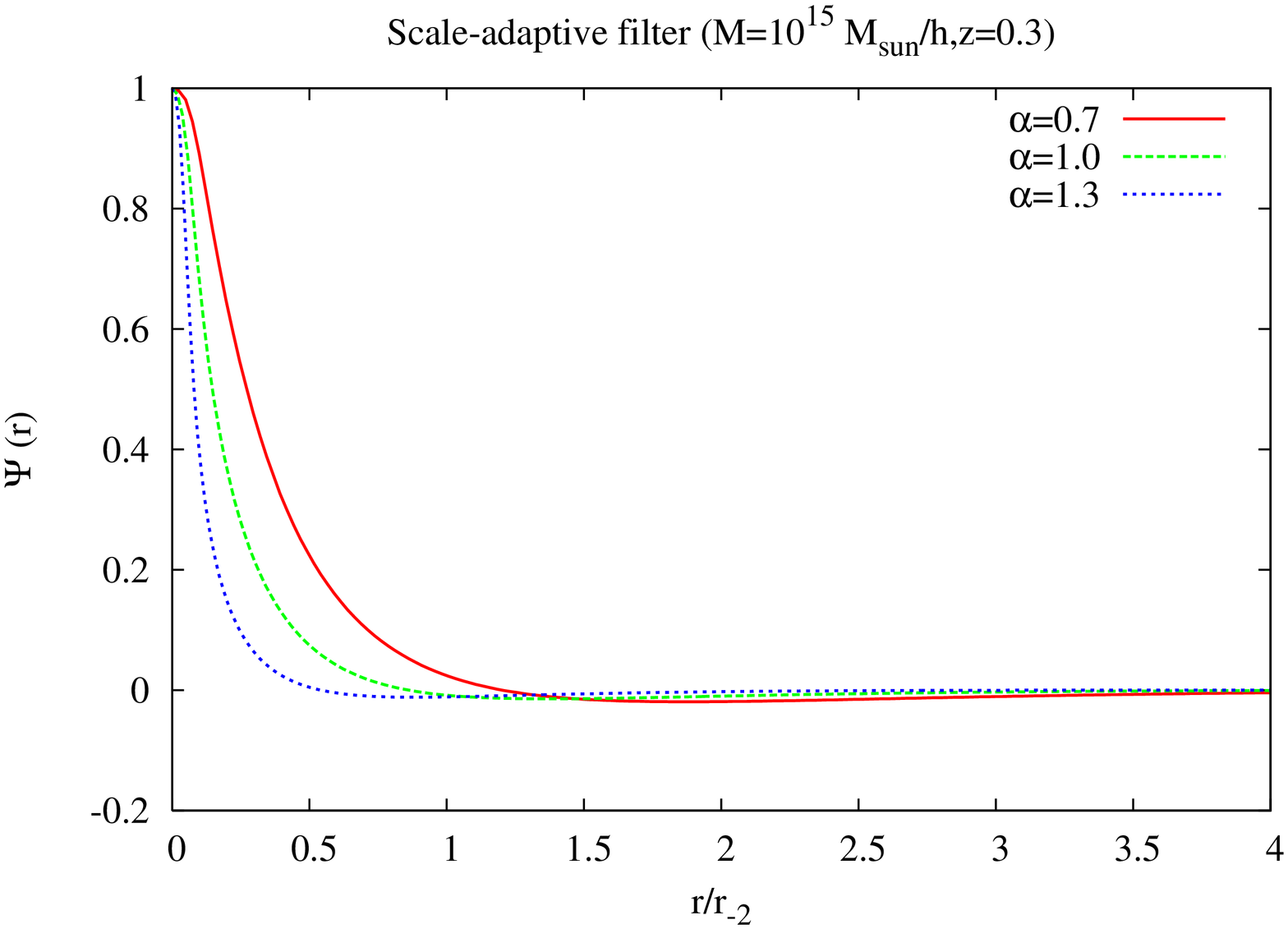}} 
    \end{tabular}
    \caption{Optimal linear filter (\textbf{left panel}) and scale-adaptive filter (\textbf{right panels}) shapes calculated for three different values of $\alpha$. All filters are normalised to unity.}\label{fig:filter_shape}
    \end{minipage}
  \end{center}
\end{figure*}

The left panel of Fig.~\ref{fig:filter_shape} shows the filter's shape calculated using three different values of the inner slope, $\alpha=0.7, 1.0, 1.3$.

\subsubsection{Dealing with non-linear signals}\label{sec:nonLinear}

The filter described in the previous section can be used to measure quantities which appear linearly in Eq.~\ref{eq:data} (e.g.~the amplitude of the shear signal). This is not the case for the inner slope $\alpha$ breaking the main assumption on which the linear filter is based on. However, if we expand the halo's shear profile around a fiducial value of the inner slope, $\alpha_0$,
\begin{equation}
\gamma(\vec{\theta},\alpha,\vec{w})=\gamma(\vec{\theta},\alpha_0,\vec{w})+\frac{\partial \gamma(\vec{\theta},\alpha,\vec{w})}{\partial \alpha}\Bigg\vert_{\alpha_0}\Delta \alpha ,\label{lin_prof}
\end{equation}
Eq.~\ref{eq:data} reads
\begin{equation}
D(\vec{\theta})-\gamma(\vec{\theta},\alpha_0,\vec{w})=\frac{\partial \gamma(\vec{\theta},\alpha,\vec{w})}{\partial \alpha}\Bigg\vert_{\alpha_0}\Delta \alpha + N(\theta), \label{eq:linder}
\end{equation}
such that $\Delta \alpha$ appears linearly and the linear filtering scheme can be applied. The shear derivative with respect to $\alpha$ plays the role of the signal shape, $\tau$, and $\Delta \alpha$ that of the amplitude $A$ to be measured. This allows the definition of the following estimator for the inner slope,
\begin{equation}
\alpha^{est} = \int \Delta\gamma(\vec{\theta},\alpha,\alpha_0,\vec{w})\Psi (\vec{\theta},\alpha_0,\vec{w})d^2\theta+\alpha_0 ,
\end{equation}\label{eq:p1_aest}
where
\begin{equation}
\Delta\gamma(\vec{\theta},\alpha,\alpha_0,\vec{w}))=\gamma(\vec{\theta},\alpha,\vec{w})-\gamma(\vec{\theta},\alpha_0,\vec{w}).
\end{equation}

\begin{figure}
 \centering
   {\psfig{figure=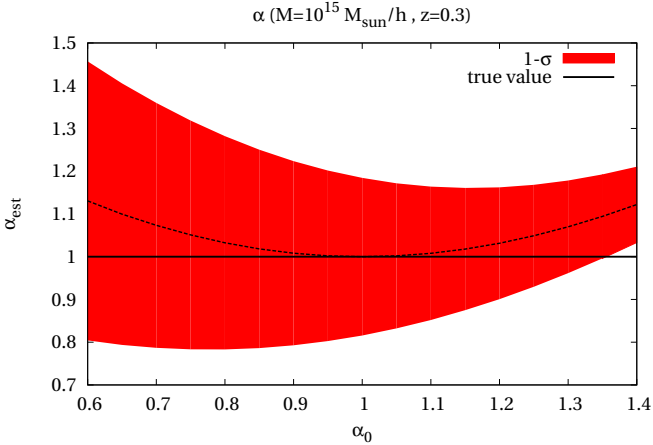,width=9cm,angle=270}}
\caption{Estimated inner slope $\alpha_{est}$ as a function of the fiducial inner slope $\alpha_0$ used in the linear filter with the 1-$\sigma$ error bars calculated via Eq.~\ref{var_flt}. The black line shows the real value of the halo's inner slope.}\label{fig:Lin_alpha}
\end{figure}

The approximation applied in Eq.~\ref{eq:linder} implies that $\alpha^{est}$ is a good estimator of the inner slope only when $\alpha_0$ is close to the real value of $\alpha$. If this is not the case, the value of the inner slope tends to be overestimated as we show in Fig.~\ref{fig:Lin_alpha}.
If a single halo is considered, the error bars associated to $\alpha_{est}$ are so large that the overestimation can be neglected for a large range of $\alpha_{0}$.  However, if several haloes are stacked, the error bars shrink and the overestimation becomes important. In order to avoid this problem, more measurements of the same halo have to be carried out sequentially: the first measurement starting with an arbitrary value of $\alpha_0$, and the second using the estimate $\alpha^{est}$ found previously as a fiducial value. We tested that, for a reasonable guess of the first fiducial value, 2-3 measurements suffice to recover the correct slope.

\subsection{Scale-adaptive filter}\label{sec:SA}

The linear expansion used in the previous section can be avoided by defining a scale-adaptive filter. Such a filter is defined similarly as the linear filter from Sect.~\ref{sec:LF} with an additional constraint on the amplitude of the signal $A_{est}$ which must be maximised when the adopted inner slope fits the data best,
\begin{equation}
\xi=\frac{\partial A_{est}}{\partial \alpha}\Bigg|_{\alpha_0}=0.
\end{equation}
The minimisation of $L=\sigma^2+\lambda_1 b+\lambda_2 \xi$ leads to the filter
\begin{equation}
\hat{\Psi}(\vec{k},\alpha)=\frac{1}{2\pi}\frac{\hat{\tau}(\vec{k},\alpha)}{P_N(k)}\frac{1}{\Delta}\Bigg[2b+c-(2a+b)\frac{d\ln\hat{\tau}(\vec{k},\alpha)}{d\ln \alpha}\Bigg] \, \label{matched_fltr}
\end{equation}
with the constants
\begin{equation}
a=\frac{1}{2\pi}\int dk k\frac{\hat{\tau}(\vec{k},\alpha)}{P_N(k)} \,
\end{equation}
\begin{equation}
b=\frac{1}{2\pi}\int dk k \frac{k}{P_N(k)}\frac{d\hat{\tau}(\vec{k},\alpha)}{d\ln \alpha} \,
\end{equation}
\begin{equation}
c=\frac{1}{2\pi}\int dk k \frac{1}{P_N(k)}\Bigg(\frac{d\hat{\tau}(\vec{k},\alpha)}{d\ln \alpha}\Bigg)^2 \,
\end{equation}
\begin{equation}
\Delta = ac-b^2\;.
\end{equation}

Its defining property is thus to maximise the signal-to-noise ratio when the correct inner slope is adopted. This implies that the inner slope can only be determined indirectly from a sequence of measurements of the shear amplitude $A_{est}$, searching for that value of $\alpha$ that maximises $A_{est}$.

The filter shape is plotted in the right panel of Fig.~\ref{fig:filter_shape}.

\subsection{Dealing with parameter degeneracies}\label{sec:degeneracy}

The two methods presented in Sects.~\ref{sec:LF} and \ref{sec:SA} assume a cluster model with known mass, redshift and concentration. In a realistic situation, we can assume to have sufficiently precise redshifts. Mass estimates would have to be obtained from optical richness, kinematics of the cluster galaxies or X-ray scaling relations. Then, estimates for the concentration could be derived from the mass-concentration relation found in numerical simulations, albeit with a considerable scatter. The concentration depends only very weakly on the mass, hence uncertainties in the mass estimate do not strongly affect the concentration estimate, and thus the mass does not need to be precisely known. However, numerical simulations suggest a log-normal distribution of the concentration around its mean with a standard deviation of $\sim0.2$, which implies that concentration parameters of real clusters can only be very poorly guessed.

Moreover, the inner slope, as the parameter we are aiming to measure, is degenerate with the concentration. In fact, it is possible to describe a halo with high central density with a large value of $\alpha$ and a small value of $c$ or vice versa, and so the problem is not well defined \citep{Wyithe01}. Thus, any attempt at measuring the profile's inner slope depends critically on the assumed halo concentration, which is uncertain in reality.

To cope with this problem, it is convenient to re-parametrise the profile accounting for this model degeneracy,
defining new parameters which can be more precisely measured. In short, the logic behind the procedure described below is as follows. In a realistic situation, we have no chance to break the degeneracy between $c$ and $\alpha$. Rather, we can rotate the parameter space such that one of its axes becomes parallel to the degeneracy direction and the other perpendicular to it. The latter will define a new parameter as a linear combination of $c$ and $\alpha$ which observations can constrain best. Comparisons with theory should then be performed on the basis of this parameter rather than through $c$ and $\alpha$ separately.

\begin{figure*}
  \begin{center}
  \begin{minipage}{180mm}
    \begin{tabular}{cc}
    \resizebox{90mm}{!}{\includegraphics[width=7.0cm,angle=270]{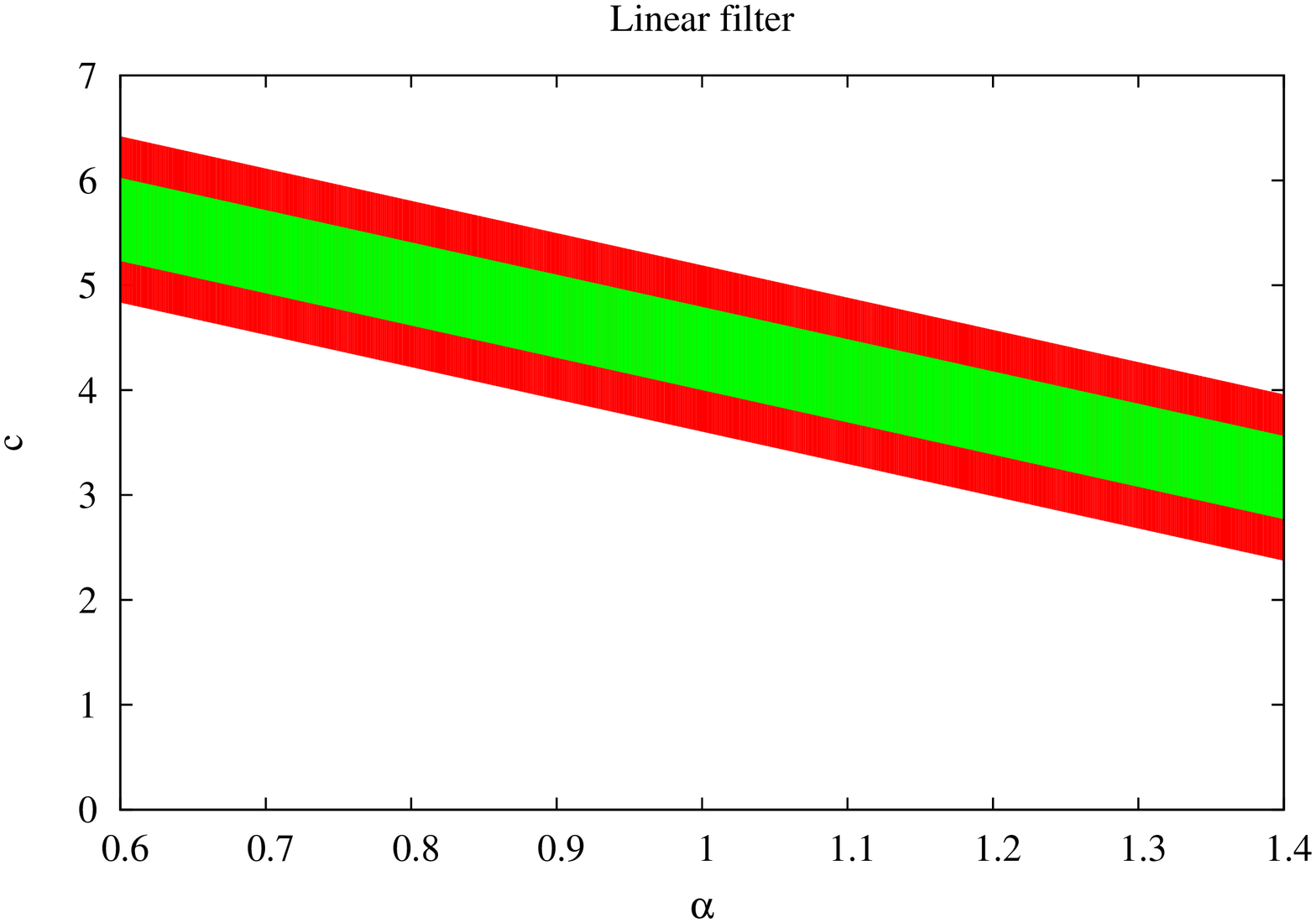}} &
    \resizebox{90mm}{!}{\includegraphics[width=7.0cm,angle=270]{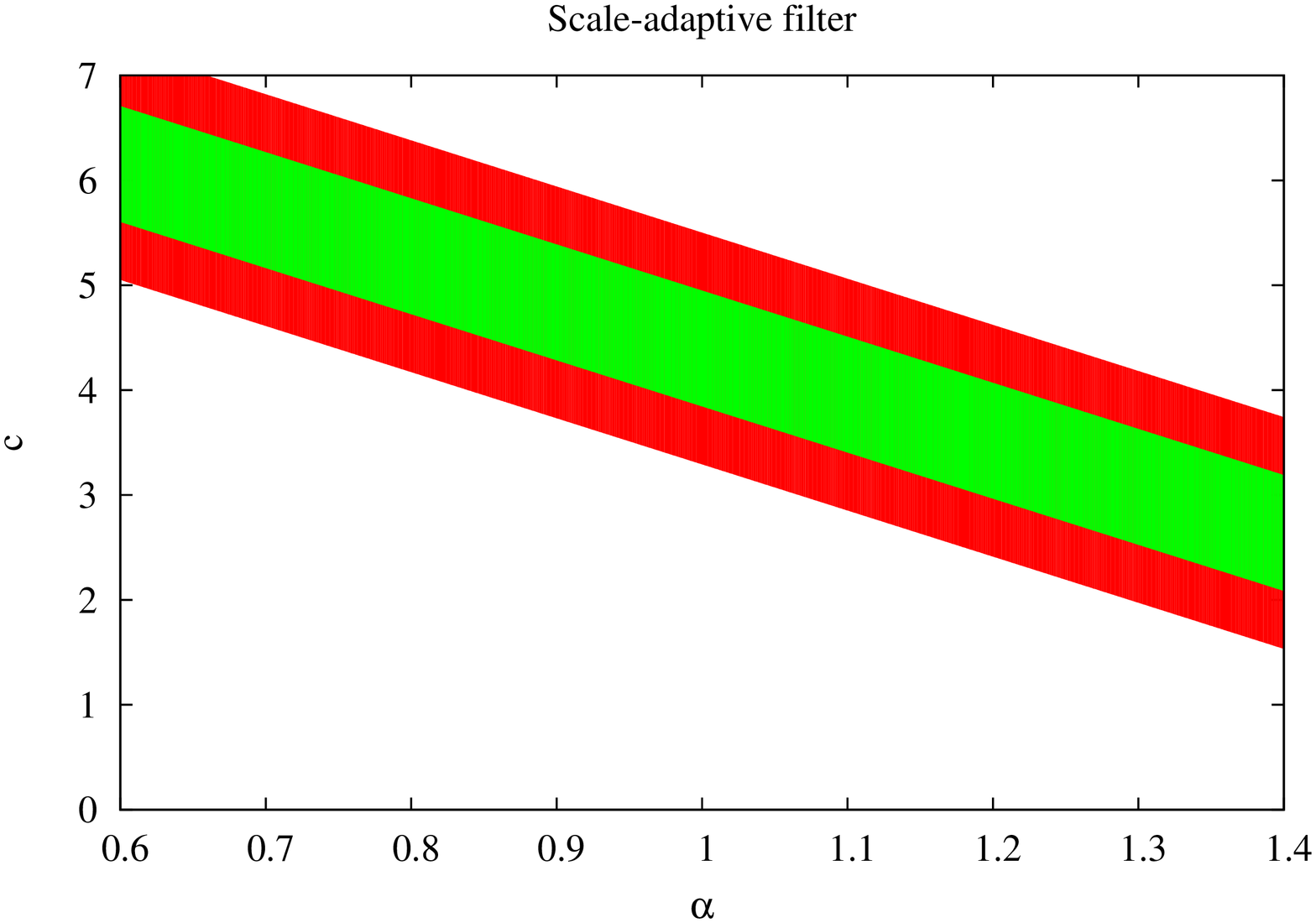}} \\
    \end{tabular}
    \caption{1-$\sigma$ and 2-$\sigma$ likelihood regions in the plane $(\alpha,c)$ computed using the linear filter (left panel) and the scale-adaptive filter (right panel) for a halo of $M=5\times 10^{14} M_{\odot}/h$ at $z=0.3$. The fiducial value is $(1.0,4.4)$.}\label{fig:Fisher}
    \end{minipage}
  \end{center}
\end{figure*}

This is achieved by a Fisher-matrix analysis. The Fisher matrix is
\begin{equation}
F_{ij}=\Bigg\langle\frac{-\partial^2 L}{\partial\pi_{i} \partial\pi_{j}}\Bigg\rangle,
\end{equation}
where $L$ is the logarithm of the likelihood function and $\vec{\pi}=(\alpha,c)$ are the free model parameters. In case of a Gaussian probability distribution, the Fisher matrix can be written as
\begin{equation}
F_{ij}=\frac{1}{2}Tr[A_{i}A_{j}+C^{-1}M_{ij}] \, \label{eq:Fisher}
\end{equation}
where $C$ is the covariance matrix, $A_i=C^{-1}C_{,i}$ and $M_{ij}=2\frac{\partial \mu}{\partial \pi_i}\frac{\partial \mu}{\partial \pi_j}$ and $\mu$ is the assumed model. Since $C$ does not depend on the inner slope and on the concentration, the first term in Eq.~\ref{eq:Fisher} vanishes. We evaluate the Fisher matrix at a fiducial point $(\alpha_0,c_0)$. In particular, we assume $\alpha_0=1$ and we calculate $c_0$ using the prescription by \citep{Eke01}. We truncate the shear profile at an inner radius $r_{min}=1/\sqrt{n_{gal}}$, which is the minimum achievable resolution for a given number density $n_{gal}$ of background galaxies and at an outer radius $r_{out}=r_{200}$. Once $r_{out}>r_s$ the Fisher matrix depends negligibly on $r_{out}$ since the derivative of the shear profile with respect to alpha is zero and the derivative with respect to the concentration is very small.

The eigenvectors $(v_1, v_2)$ and $(v_3, v_4)$, of the Fisher matrix, determining the directions of largest and smallest degeneracy between the parameters $\alpha$ and $c$, define a rotation of the parameter space and thus two new parameters
\begin{eqnarray}
P_1=v_1\alpha + v_2c \\ 
P_2=v_3\alpha + v_4c \,
\end{eqnarray} 
which are linear combinations of $\alpha$ and $c$. The two new parameters are those which can be constrained best and worst, respectively, given the model adopted in the Fisher-matrix estimate.

If the linear filter is used to measure the inner slope, the model $\mu$ is
\begin{equation}
\mu=\alpha_{est}(\alpha,c)=\int [\gamma(\alpha,c)-\gamma(\alpha_0,c_0)]\Psi(\alpha_0,c_0)d^2x	
\end{equation}
and thus
\begin{equation}
\frac{\partial \mu}{\partial \pi_i}=\int \frac{\partial \gamma(\vec{x},\vec{\pi})}{\partial \pi_i}\Psi(\vec{x},\vec{\pi}_0)d^2x.
\end{equation}
The covariance matrix reduces in this case to the variance of the measurement obtained from Eq.~\ref{var_flt}.

Note that the Fisher matrix defined above is singular, i.e.~its determinant vanishes. The errors on the new parameters are given by $1/\sqrt{\lambda_i}$, where $\lambda_i$ are the eigenvalues of the Fisher matrix. Since one of them is vanishing the error on one parameter (taken to be $P_2$) is infinite. This means that the likelihood region in the plane $(\alpha,c)$ is an ellipse infinitely elongated in the degeneracy direction. This is because there is more than one way of fitting a single data set ($\Delta \alpha$) by varying the two parameters. In the right panel of Fig.~\ref{fig:Fisher} we show the result for a halo of $M=5\times 10^{14} M_{\odot}/h$ at redshift $z=0.3$ with concentration $c=4.4$. The corresponding eigenvector components are $v_1=v_4=0.95$ and $v_2=-v_3=0.30$.

When the scale-adaptive filter is used, the measurable quantity is the shear amplitude
\begin{equation}
A(\alpha,c)=\int D(\alpha_H,c_H;\vec{\theta})\Psi(\alpha,c_0;\vec{\theta}) d^2\theta,
\end{equation}
and the value of the inner slope ($\alpha_{est}$) is then estimated looking for the value of $\alpha$ maximising the amplitude. It is clear that it depends only on the halo's concentration $c_0$ assumed in the filter. To find the degeneracy direction between the inner slope and the concentration in this case, we analyse the relation between $\alpha_{est}$ and $c$ around a fiducial point in the ($\alpha,c$) plane. The result is shown in the left panel of Fig.~\ref{fig:Fisher} for the same halo as considered before. Here, too, we define two new parameters $P_1=0.97\alpha+0.22c$ and $P_2=-0.22\alpha+0.97c$. In this case, the error cannot be calculated analytically since the measurement is indirect. Instead, we have performed a Monte-Carlo simulation (see Sect.~\ref{sec:uncert}). 

Since the shapes of the filters are different, so are the degeneracy directions we find.

\begin{figure}
\centering
{\psfig{figure=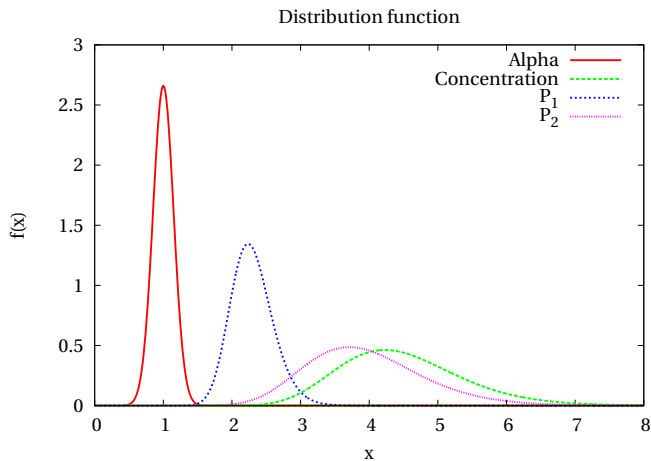,width=9cm,angle=270}}
\caption{Probability distributions for $\alpha$, $c$, $P_1$ and $P_2$. A normal distribution with $\sigma_{\alpha} = 0.15$ for the inner slope \citep{Diemand04} and a log-normal distribution with $\sigma_{c}=0.2$ for the concentration \citep{Bullock01b} are assumed. The probability distribution for $P_1$ is given by the convolution of the probability distribution of $\alpha$ and $c$, while the probability distribution for $P_2$ is given by their cross correlation. }\label{fig:distribution}
\end{figure}

The probability distributions of $P_1$ and $P_2$ can be found convolving the probability distributions of the concentration and the inner slope. Using the degeneracy direction found for the linear filter and assuming a log-normal distribution for the concentration with $\sigma_c = 0.2$ \citep{Bullock01b} and a Gaussian distribution for the inner slope with $\sigma_{\alpha}=0.15$ \citep{Diemand04}, we find that both probability distributions of $P_1$ and $P_2$ can be approximated as log-normal distributions with standard deviations $\sigma_{P_1}=0.29$ and $\sigma_{P_2 }=0.33$ respectively, as shown in Fig.~\ref{fig:distribution}.

\section{Method uncertainties}\label{sec:uncert}

Here, we discuss in detail possible error sources affecting the measurement of the inner slope using the methods described in Sects.~\ref{sec:LF} and \ref{sec:SA}. We will show the error calculation for a halo of $M=5\times 10^{14} M_{\odot}/h$, $z=0.3$, $c=4.4$.

The statistical uncertainties arising from the data noise component $N$ are given by the intrinsic ellipticity of the background galaxies, their finite number and from the contamination due to the intervening large-scale structures. The filters we have defined minimise these uncertainties. They are quantified by Eq.~\ref{var_flt} for the linear filter and by a Monte-Carlo analysis for the scale-adaptive filter since in this case $\alpha$ is measured indirectly by estimating the location of the maximum in the estimated signal, and an analytical computation of its variance is impossible.

 The Monte-Carlo analysis has been performed generating 1000 realisations of a shear catalogue using randomly distributed background galaxies with a density $n_{gal}=30/arcmin^2$, placed at redshift $z_s=1.0$, on a $0.01$ degree field. The halo has been placed in the field center. The noise due to the intrinsic galaxy ellipticities ($\sigma_{\epsilon}=0.3$) and the lensing effect due to the intervening large-scale structure have been added. The latter noise is calculated assuming that the large-scale structure can be described by a Gaussian random field with a power spectrum determined by the linear theory of structure growth.
 
\begin{figure}
\begin{center}
\hbox{
\psfig{figure=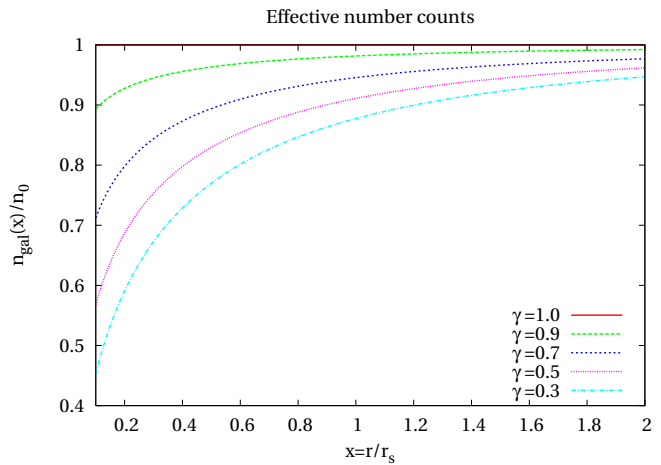,width=9cm,angle=270}
}
   \caption{Magnification bias expected for an halo of $5\times 10^{14} M_{\odot}/h $ at redshfit $z=0.3$ lensing galaxies at $z=1.0$. $\gamma$ is the exponent of the power low in equation \ref{eq:numCounts}.}\label{fig:magBias}
   \end{center}
\end{figure}

We assume in our analysis that the magnification bias can be neglected, allowing us to leave the effective number $n_{gal}$ of available galaxies unchanged. This is justified only if the slope $\gamma$ of the flux distribution of faint galaxies
 \begin{equation}
 n_{0}(>S)=aS^{-\gamma} \label{eq:numCounts}
 \end{equation}
is unity as discussed by \cite{Bartelmann01}. The effective number of galaxies $n_{eff}$ scales with $\gamma$ as
  \begin{equation}
        \frac{n_{eff}(>S)}{n_{gal}(>S)}=\mu^{\gamma -1}
  \end{equation}
where $\mu$ is the magnification. Specifically, $n_{eff}$ is lowered by at most 40\% compared to $n_{gal}$ near $r=0.2r_s$ if $\gamma$ is $0.5$, as shown in Fig.\ref{fig:magBias}. For galaxies in the Hubble Ultra Deep Field \citep{Beckwith06} we estimate $\gamma \simeq 0.8$ causing a magnfication bias of around $10 \%$.

\begin{figure*}
  \begin{center}
	\hbox{
	\psfig{figure=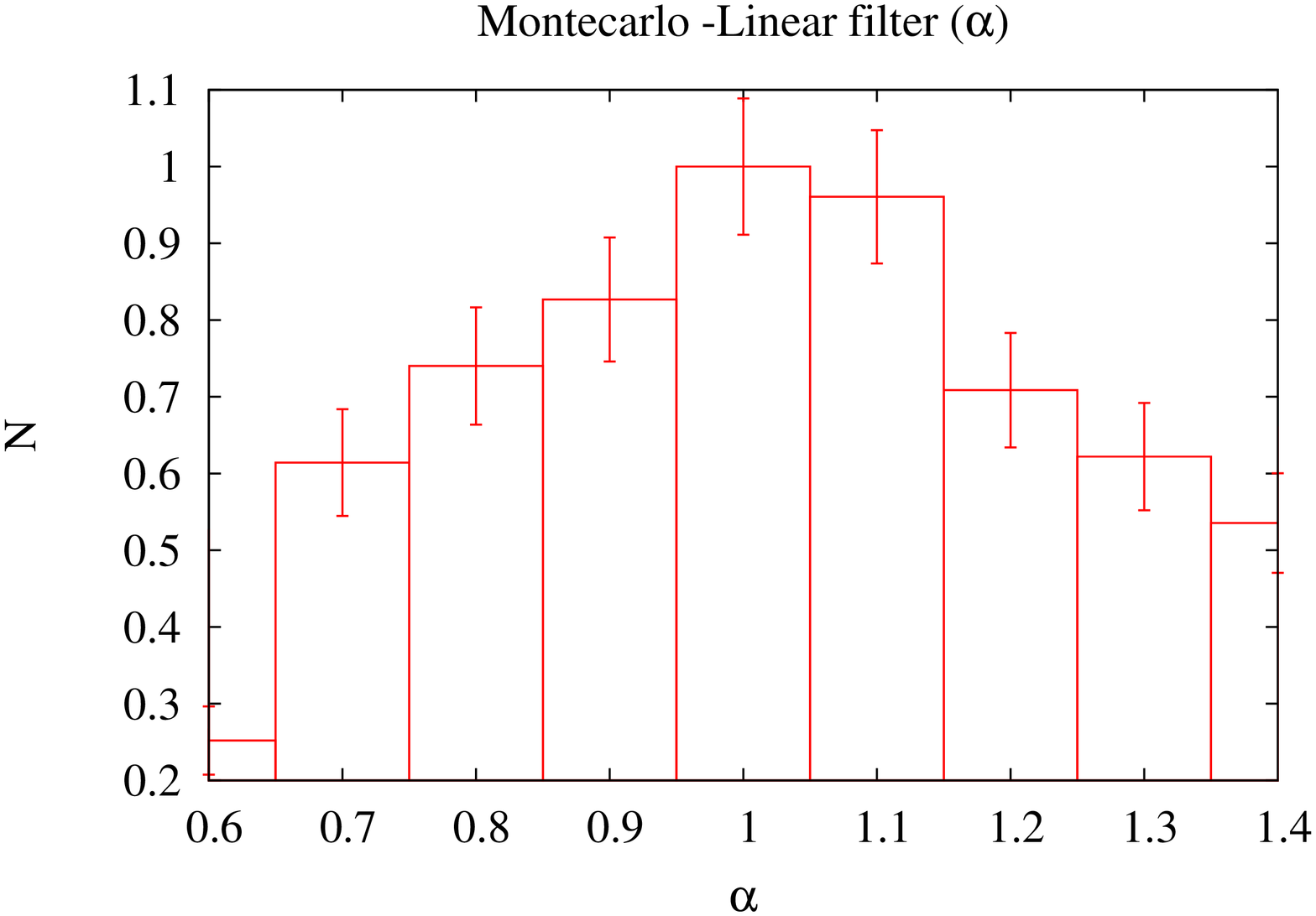,width=9cm,angle=270}
	\psfig{figure=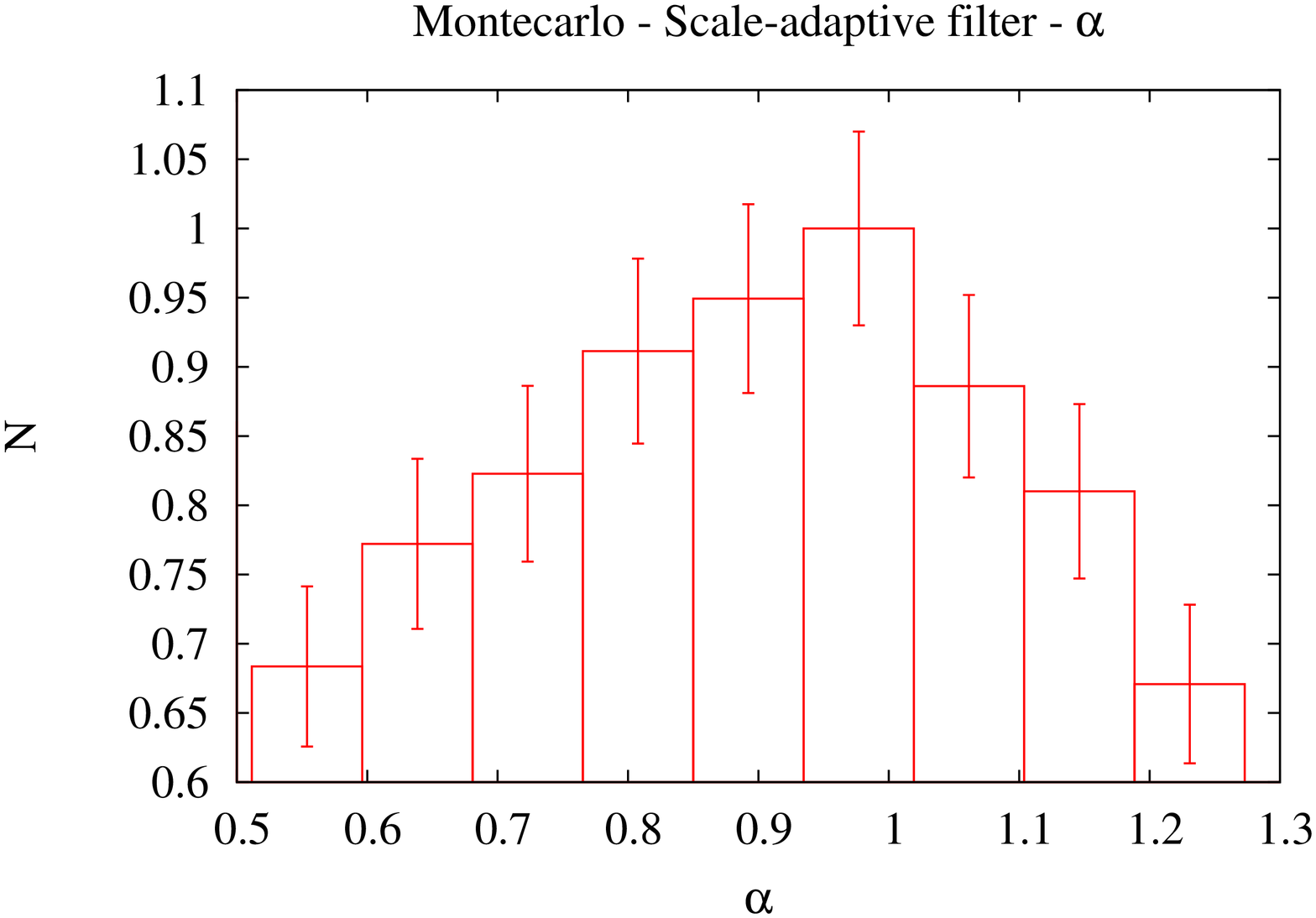,width=9cm,angle=270}
     }
  \end{center}   
    \caption{Normalised distributions of the value of the inner slope computed using the linear filter (\textbf{left panel}) and scale-adaptive filter (\textbf{right panel}). A Gaussian distribution has been assumed for the mass and the redshift, while a log-normal distribution has been adopted for the concentration.}\label{fig:Montecarlo_lin_alpha}
\end{figure*}

\begin{figure*}
  \begin{center}
  \hbox{
	\psfig{figure=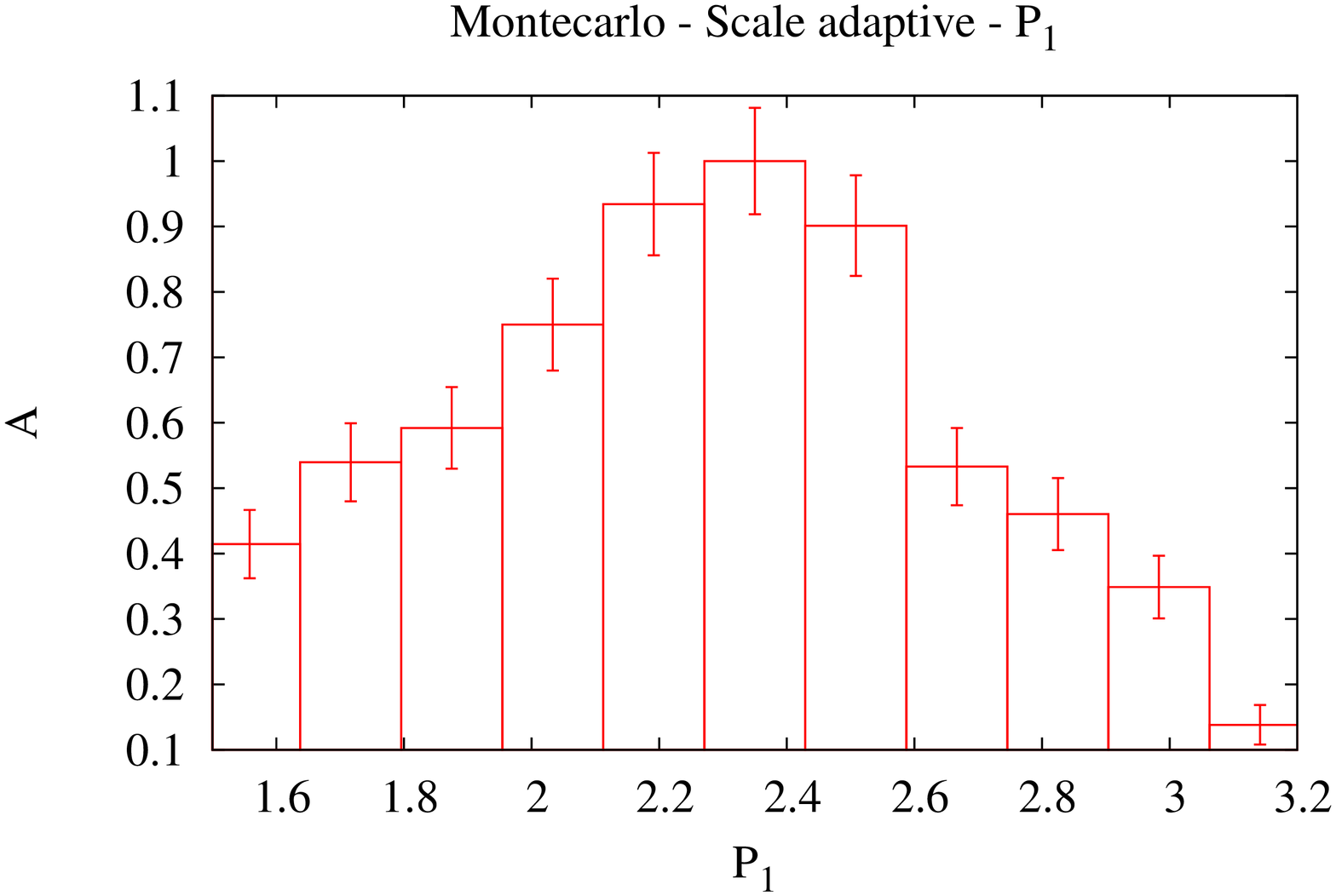,width=9cm,angle=270}
	\psfig{figure=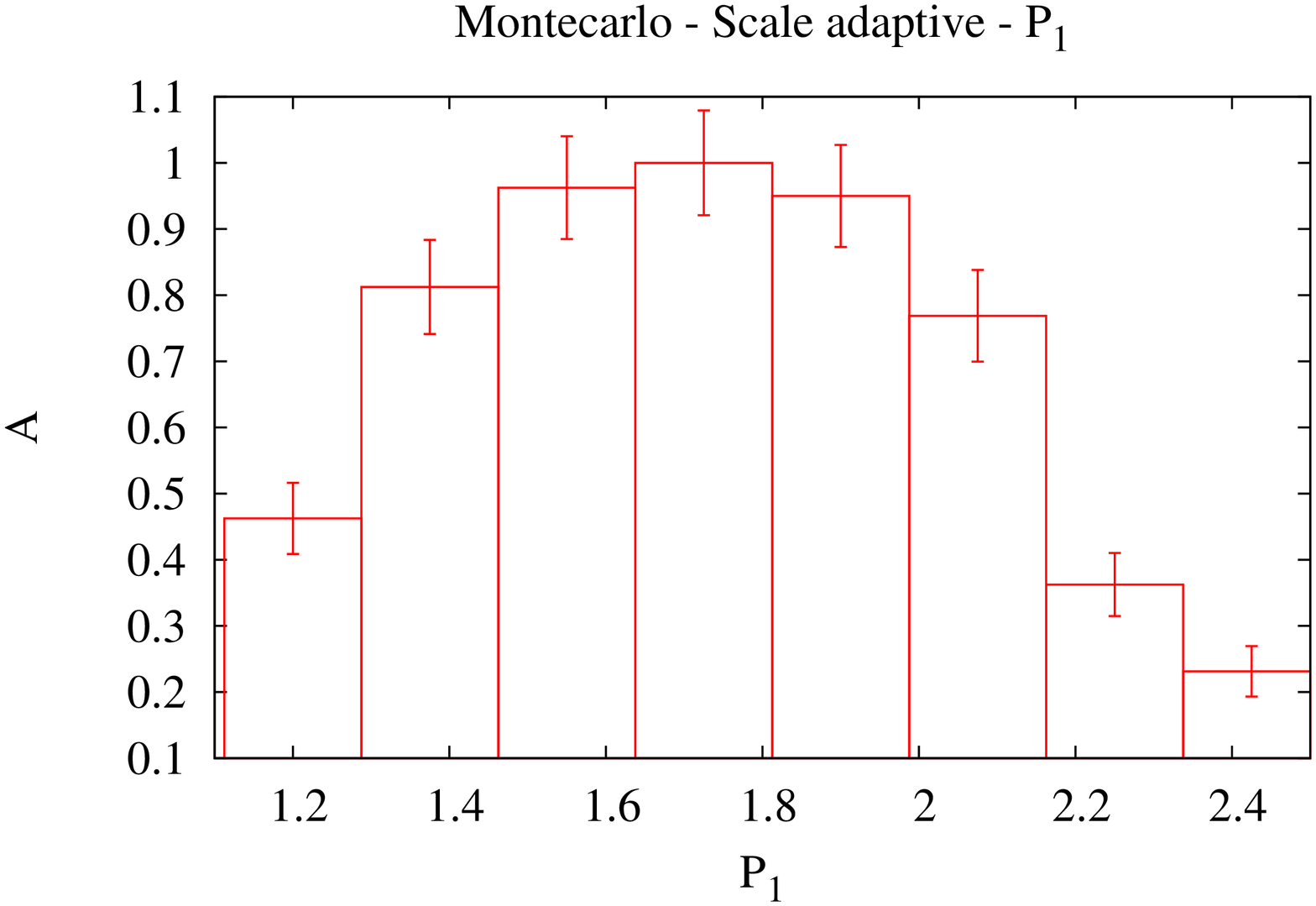,width=9cm,angle=270}
     }
      \end{center}
    \caption{As Fig.~\ref{fig:Montecarlo_lin_alpha}, but using the new parameters $P_1$ and $P_2$ instead of $\alpha$ and $c$.}\label{fig:Montecarlo_p1}
\end{figure*}

\begin{table*}
\centering
   \caption{Statistical errors in the parameters measurement for a halo of $5\times10^{14} M_{\odot}/h$ at redshift $z=0.3$. In the \textit{first column} we indicate the used filter, in the \textit{second column} the parameter we constrain and in the \textit{third column} its fiducial value. In the \textit{fourth column} are shown the expected errors assuming randomly distributed background galaxies with intrinsic ellipticity $\sigma_{\epsilon}=0.3$ and random noise due to the large-scale structures. The errors presented in the  \textit{fifth column} take also into account Gaussian errors in the halo mass and redshift with standard deviations $\sigma_M = 1.5\times 10^{14}$ and $\sigma_z=0.03$, respectively, and a log-normal distribution for the concentration with standard deviation $\sigma_c =0.2$. When $P_1$ is estimated the probability distribution of $P_2$ is calculated from the probability distribution of the concentration assuming a Gaussian probability distribution for the inner slope with \ensuremath{\protect\sigma_{\protect\alpha}=0.15}. In the \textit{sixth column} we show the percentage error on the parameter estimation.}
\begin{tabular}{|r|r|r|r|r|r|}
  \hline
  Filter&Parameter&Fiducial value&$\sigma$ (stat.)&$\sigma$ (stat.+model)&Percentage error\\
  \hline
  $SAF$&$\alpha$&$1.00$&$0.19$&$0.26$&$0.26$\\
  $SAF$&$P1$&$1.95$&$0.21$&$0.29$&$0.15$\\
  $LF$&$\alpha$&$1.00$&$0.14$&$0.28$&$0.28$\\
  $LF$&$P_1$&$2.31$&$0.15$&$0.30$&$0.13$\\
  \hline
\end{tabular} 
\end{table*}

For each realisation we use Eq.~\ref{aest_fltr} to estimate the shear amplitude in the position corresponding to the halo's center using filters initialised with an inner slope in the range $[0.6-1.4]$. The estimated inner slope value is then defined as the value of $\alpha$ giving the maximum value of the shear amplitude. We finally calculate their distribution and the dispersion around the mean value (the results are summarised in the fourth column of Table 1).

We find that the standard deviation associated with the inner slope, measured by the scale-adaptive filter, is 0.19. The analytical calculation done for the linear filter gives a value of 0.14.

\begin{figure}
\begin{center}
\hbox{
\psfig{figure=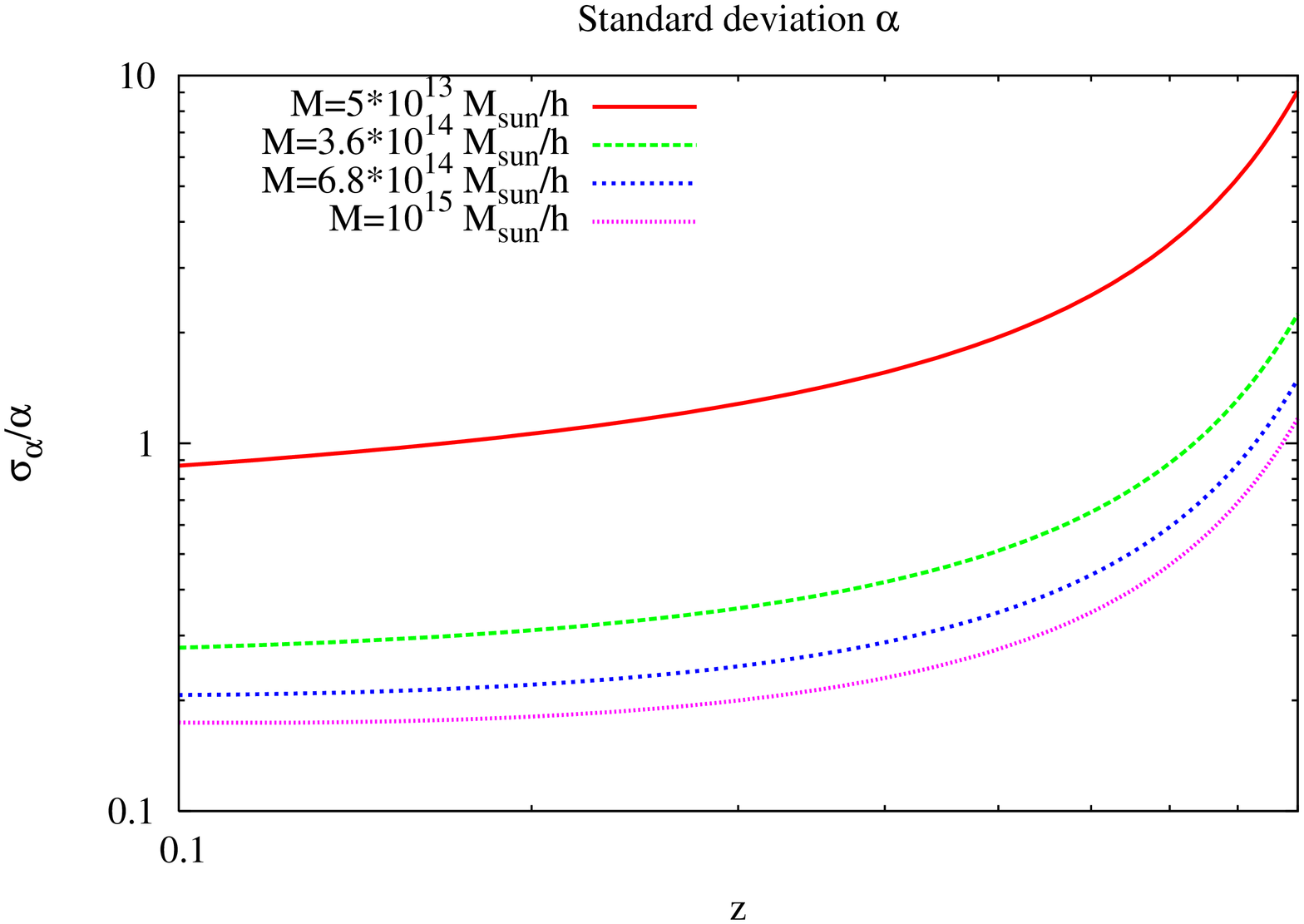,width=9cm,angle=270}
 }
\caption{Standard deviation for $\alpha$ as a function of the halo's redshift and mass.}\label{fig:var_alpha_lin}
 \end{center}
\end{figure}

The same calculation has been done considering haloes of different masses and at different redshifts. As shown in Fig.~\ref{fig:var_alpha_lin}, the standard deviation increases with respect to the redshift and decreases when the mass is increasing. In particular for a halo placed at intermediate redshift between the background sources and the observer, the standard deviation varies in the range $[0.2-1.0]$ for a mass range $[10^{15}-5\times10^{13}]$.

The preceding calculations show that errors on the inner slope due to intrinsic ellipticities of background galaxies and due to contamination by large-scale structures are large when computed for a single halo. However stacking a large number of haloes (10-100), it is possible to measure an average value of $\alpha$ with a few percent accuracy.

A more accurate error evaluation has to consider also the scatter around the fiducial value of the halo's mass, redshift and concentration used in the filter definition. For both methods, we perform a Monte-Carlo simulation, following the procedure described above, assuming a Gaussian distribution for the halo mass ($\sigma_{M}=1.5\times 10^{14}$) and redshift ($\sigma_{z}=0.03$) and a log-normal distribution for the concentration ($\sigma_{c}=0.2$) following numerical simulations (citation). The result is shown in Fig.~\ref{fig:Montecarlo_lin_alpha}.

One critical point that we have avoided so far concerns the choice of the fiducial values for the haloes parameter. We discuss this point in the following section.

\subsection{Model sensitivity}\label{sec:mod_sens}

\begin{figure*}
  \begin{center}
  \begin{minipage}{160mm}
    \begin{tabular}{cc}
    \resizebox{70mm}{!}{\includegraphics[width=7.0cm,angle=270]{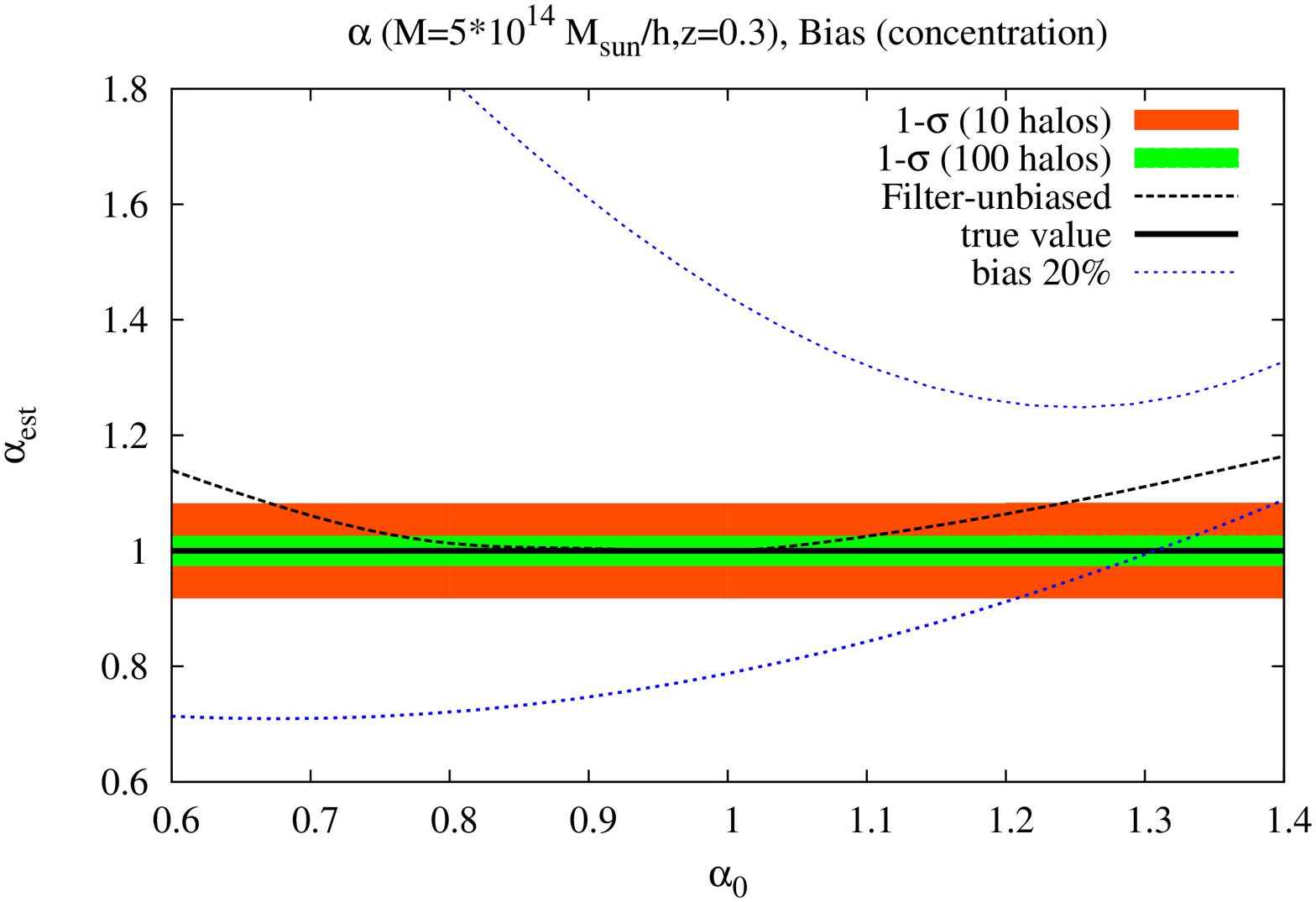}} &
    \resizebox{70mm}{!}{\includegraphics[width=7.0cm,angle=270]{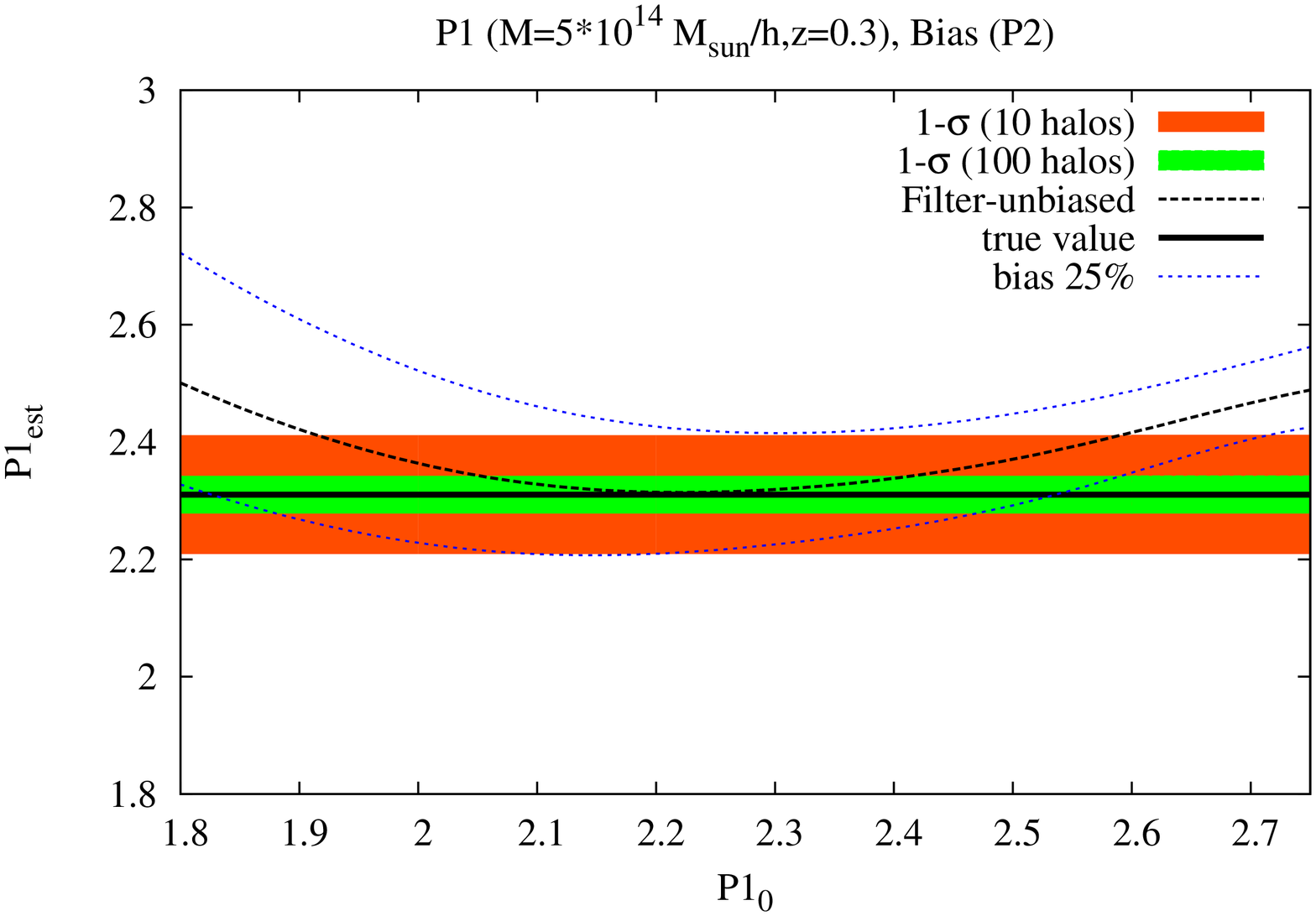}} \\
    \resizebox{70mm}{!}{\includegraphics[width=7.0cm,angle=270]{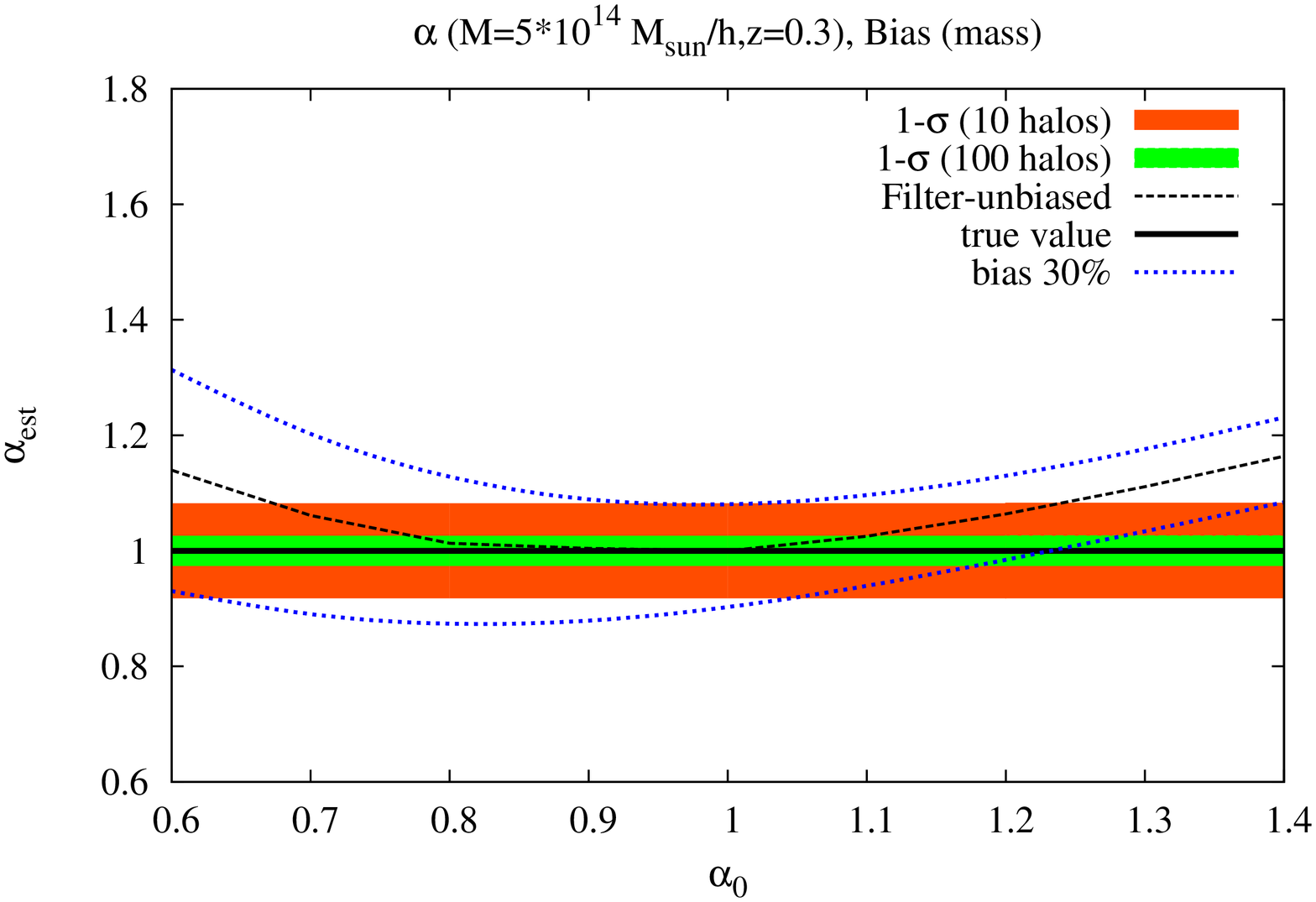}} &
    \resizebox{70mm}{!}{\includegraphics[width=7.0cm,angle=270]{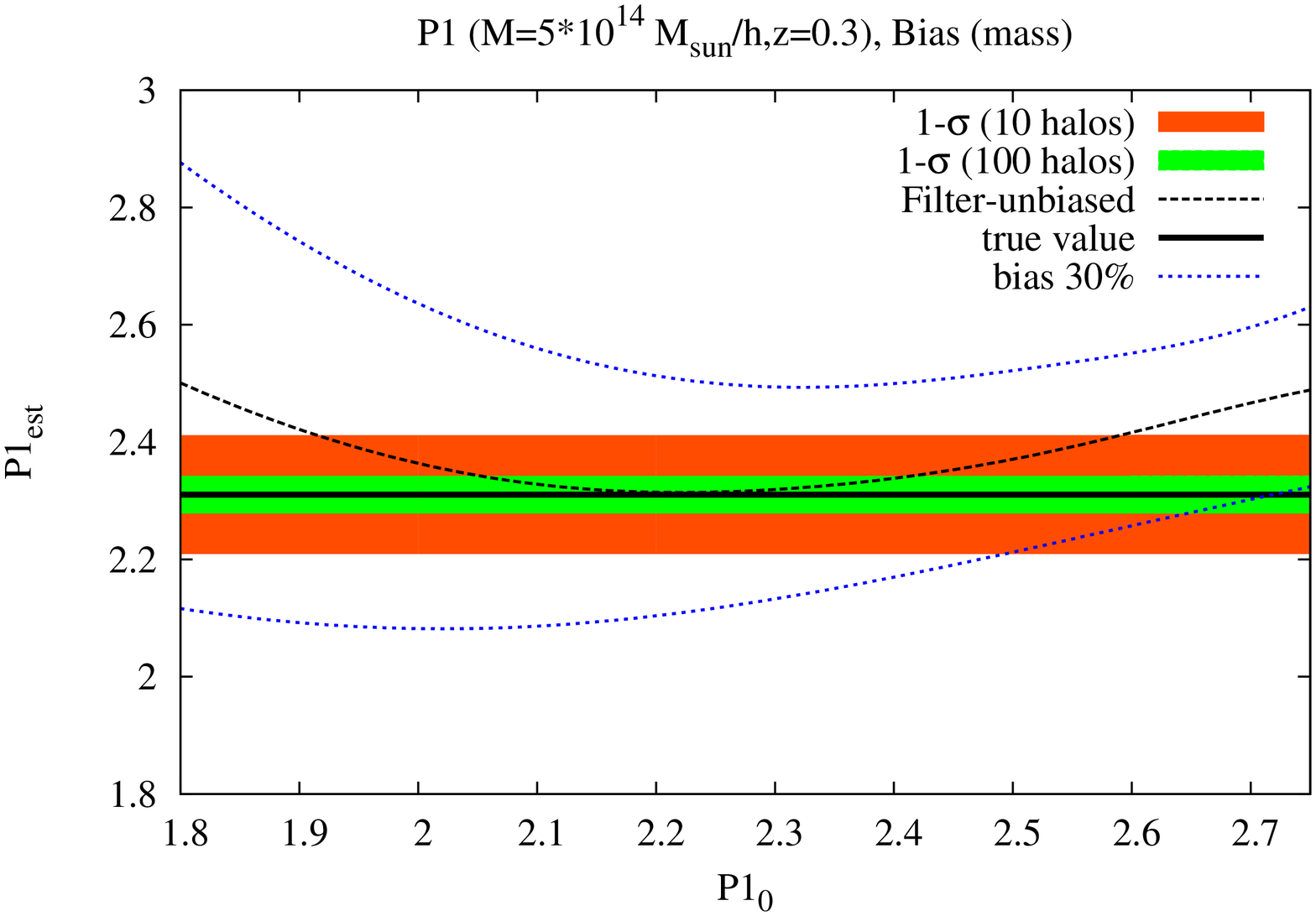}} \\
    \resizebox{70mm}{!}{\includegraphics[width=7.0cm,angle=270]{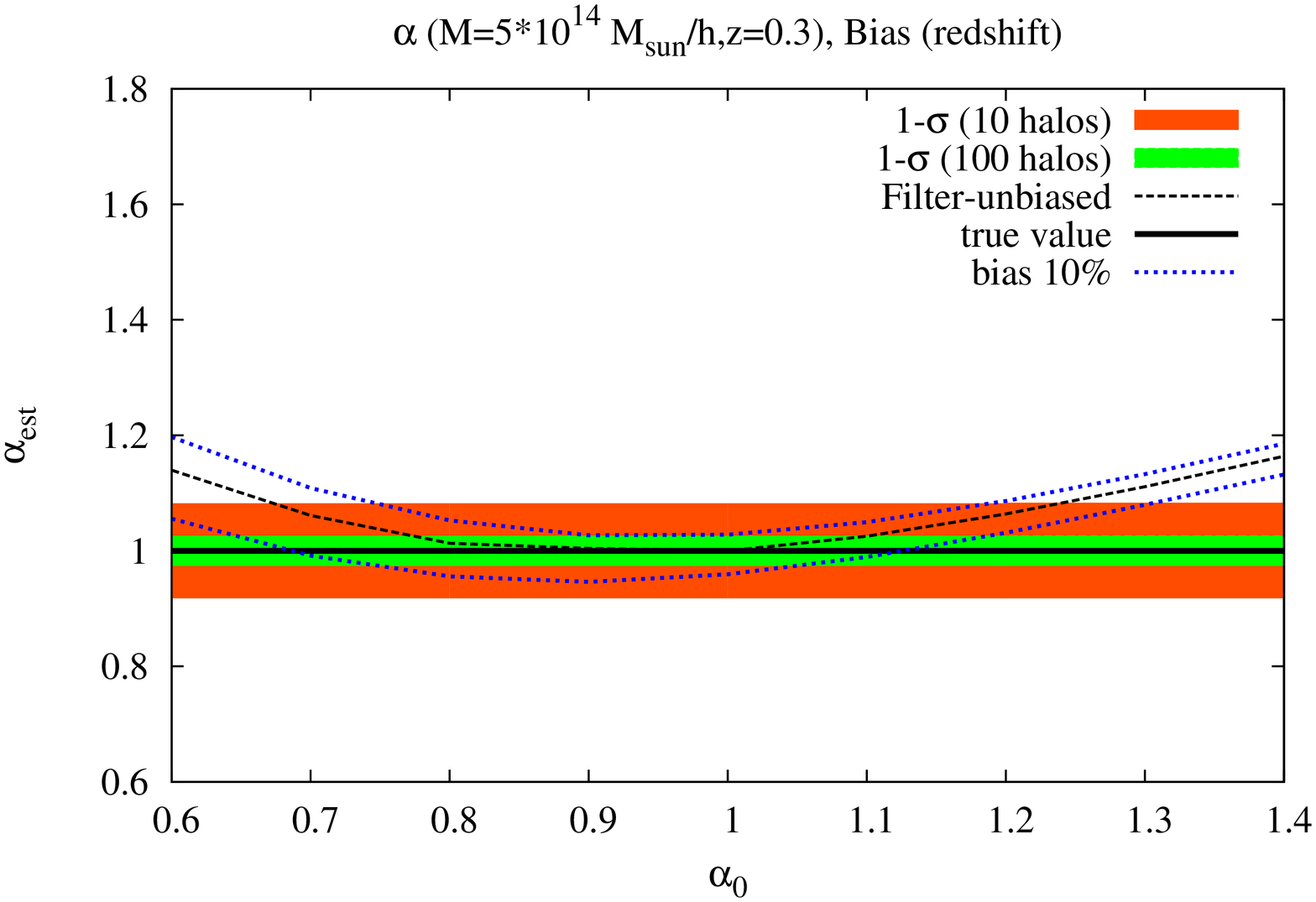}} &
    \resizebox{70mm}{!}{\includegraphics[width=7.0cm,angle=270]{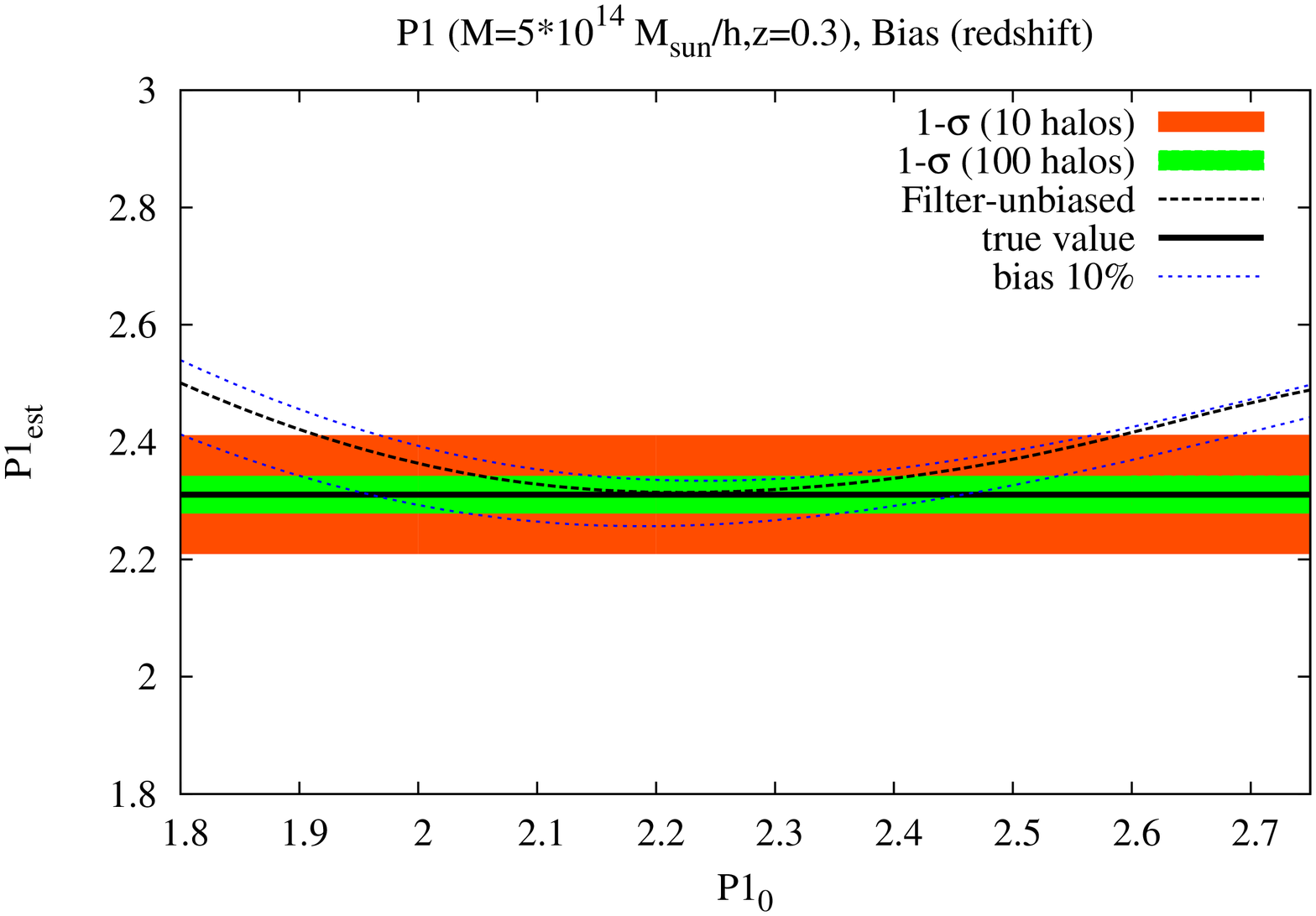}} \\
    \end{tabular}
    \caption{\textbf{Left panels:} Estimated inner slope of the halo ($\alpha_{est}$) as a function of the fiducial inner slope used in the filter ($\alpha_{0}$) with the $1-\sigma$ error calculated for 10 (orange) or 100 (green line) haloes using the Monte-Carlo simulations described in Sect.~\ref{sec:uncert}. The black line shows the real value of the halo's inner slope. The \emph{first panel} shows the bias caused by a fiducial concentration 20 \% larger or smaller than the real concentration. The \emph{second panel} shows the bias induced by a 50~\% difference between the fiducial and the real halo's mass, while the \emph{third panel} shows the bias caused by a difference of 10~\% between the fiducial and the real halo's redshift. \textbf{Right panels:} As the left panels, but for the new pair of parameters $P_1$ and $P_2$. }\label{fig:alpha_lin}
    \end{minipage}
  \end{center}
\end{figure*}

\begin{figure*}
  \begin{center}
  \begin{minipage}{160mm}
   \begin{tabular}{cc}
      \resizebox{70mm}{!}{\includegraphics[width=7.0cm,angle=270]{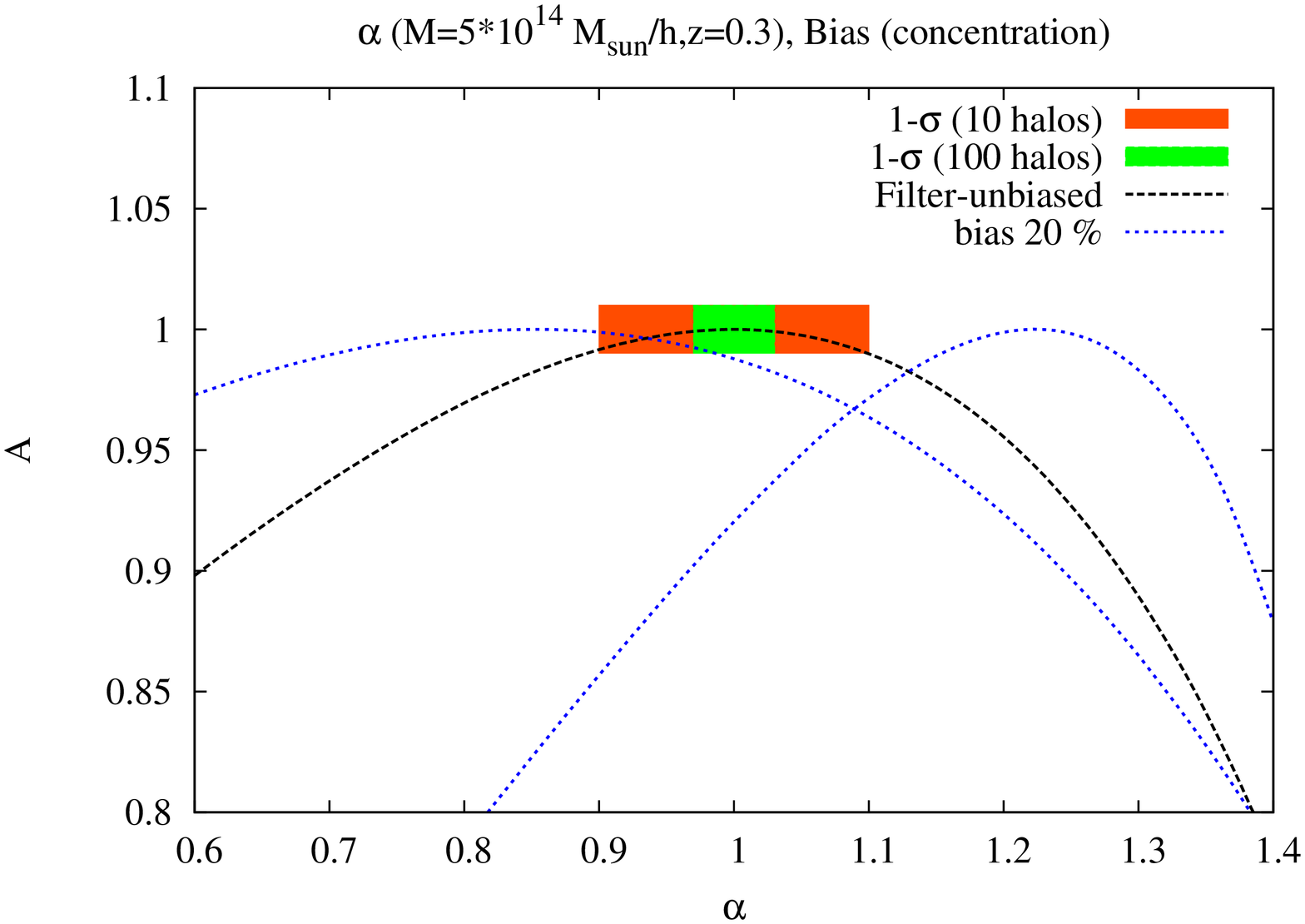}} &
      \resizebox{70mm}{!}{\includegraphics[width=7.0cm,angle=270]{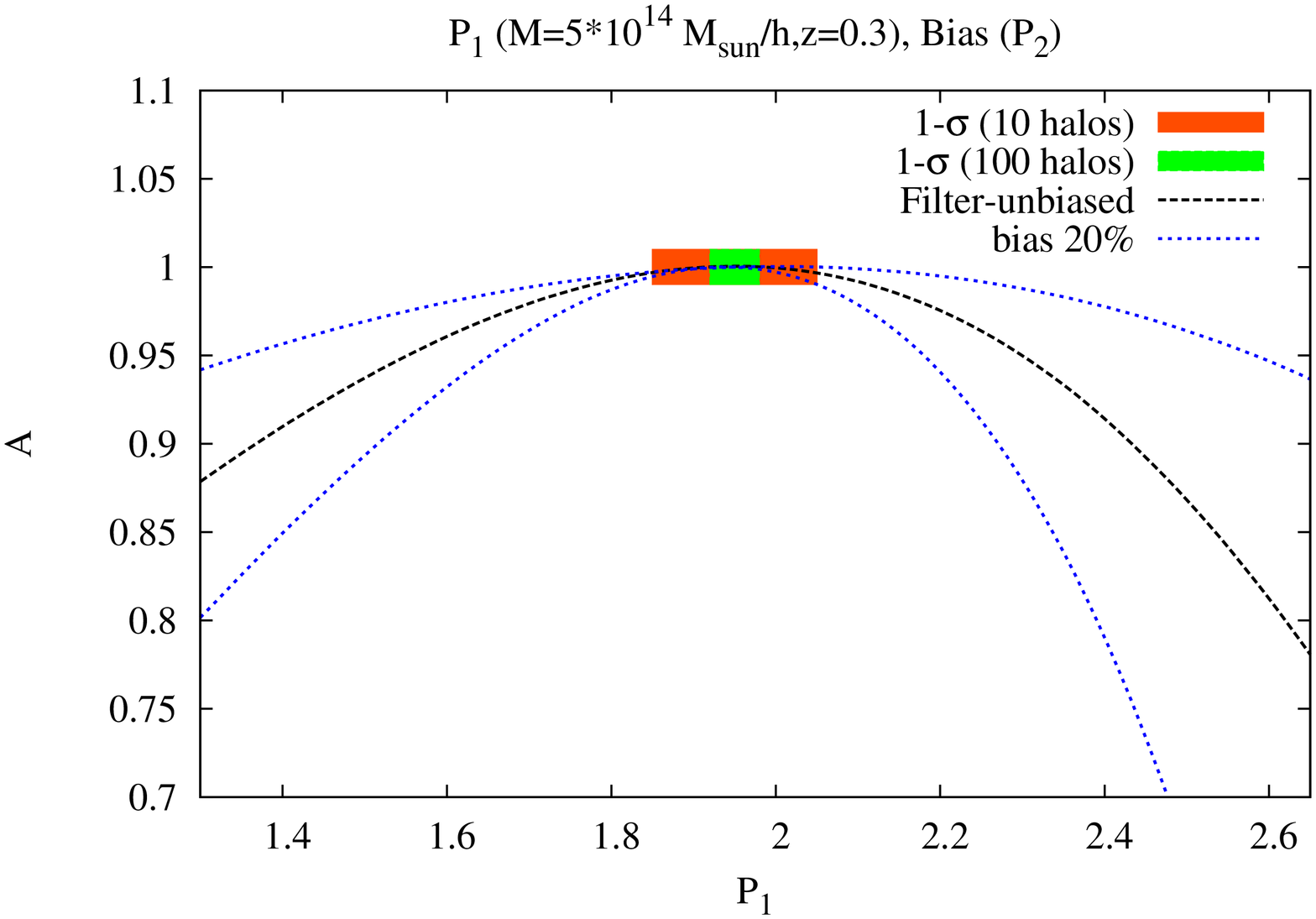}} \\
     \resizebox{70mm}{!}{\includegraphics[width=7.0cm,angle=270]{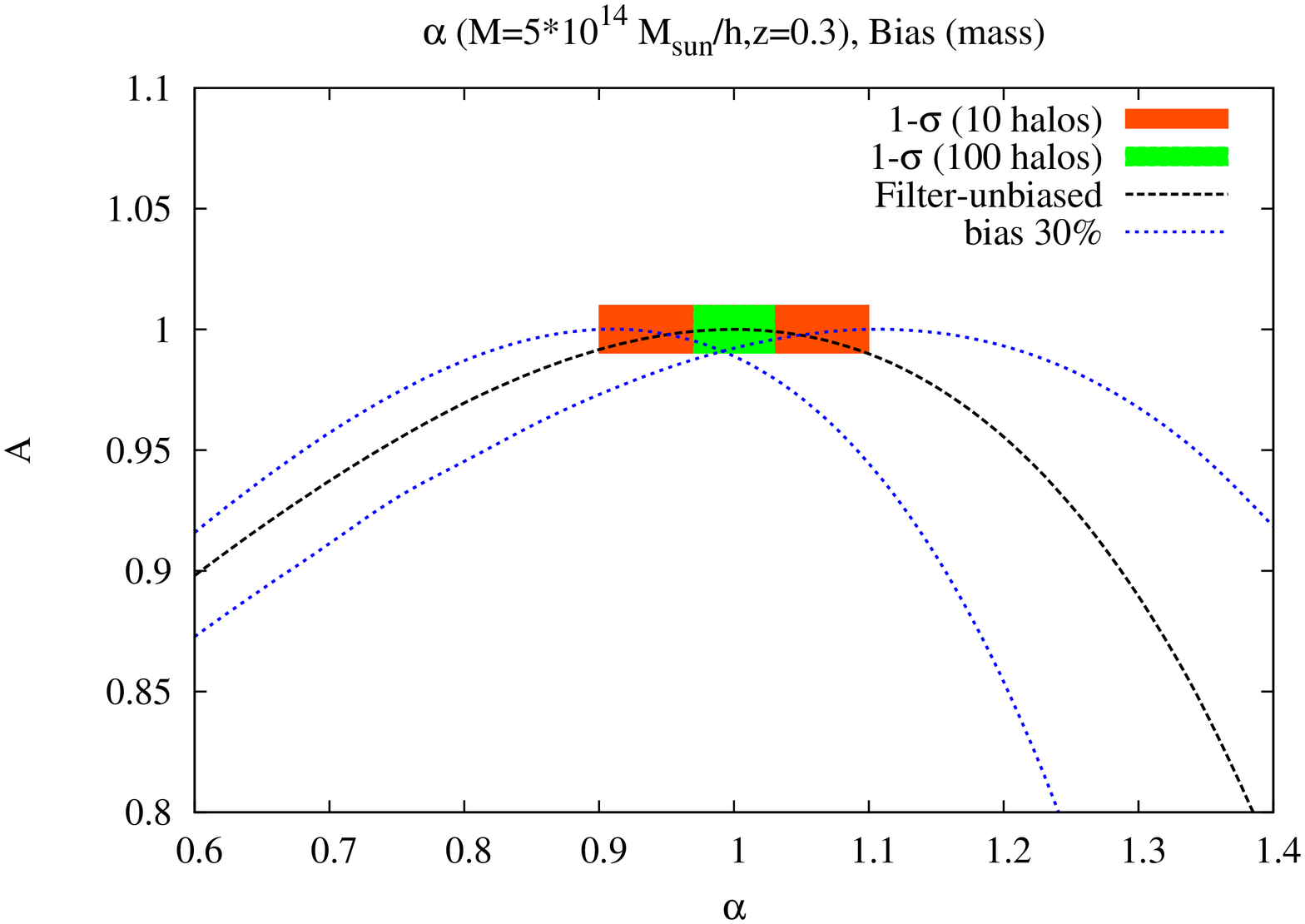}} &
      \resizebox{70mm}{!}{\includegraphics[width=7.0cm,angle=270]{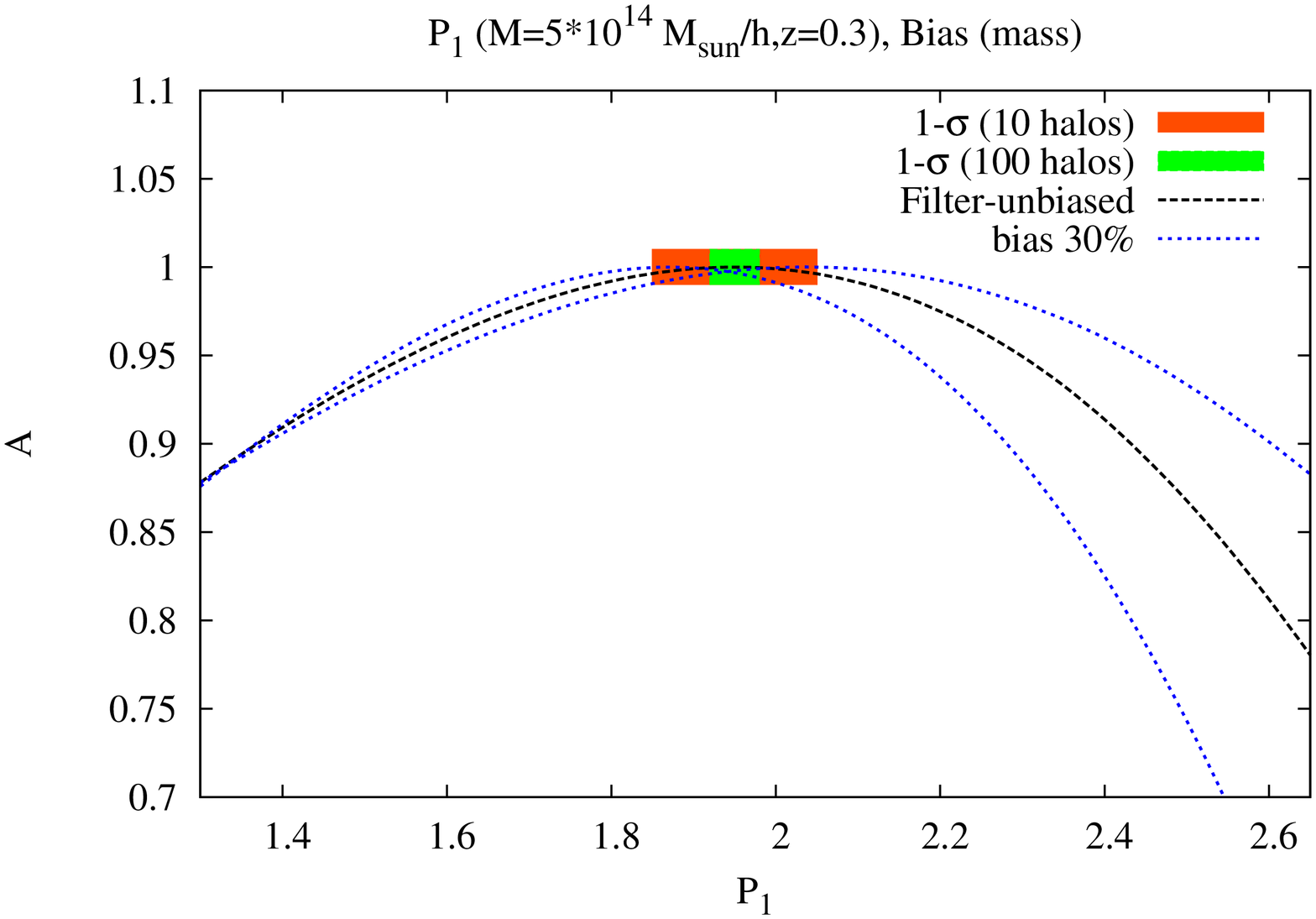}} \\
      \resizebox{70mm}{!}{\includegraphics[width=7.0cm,angle=270]{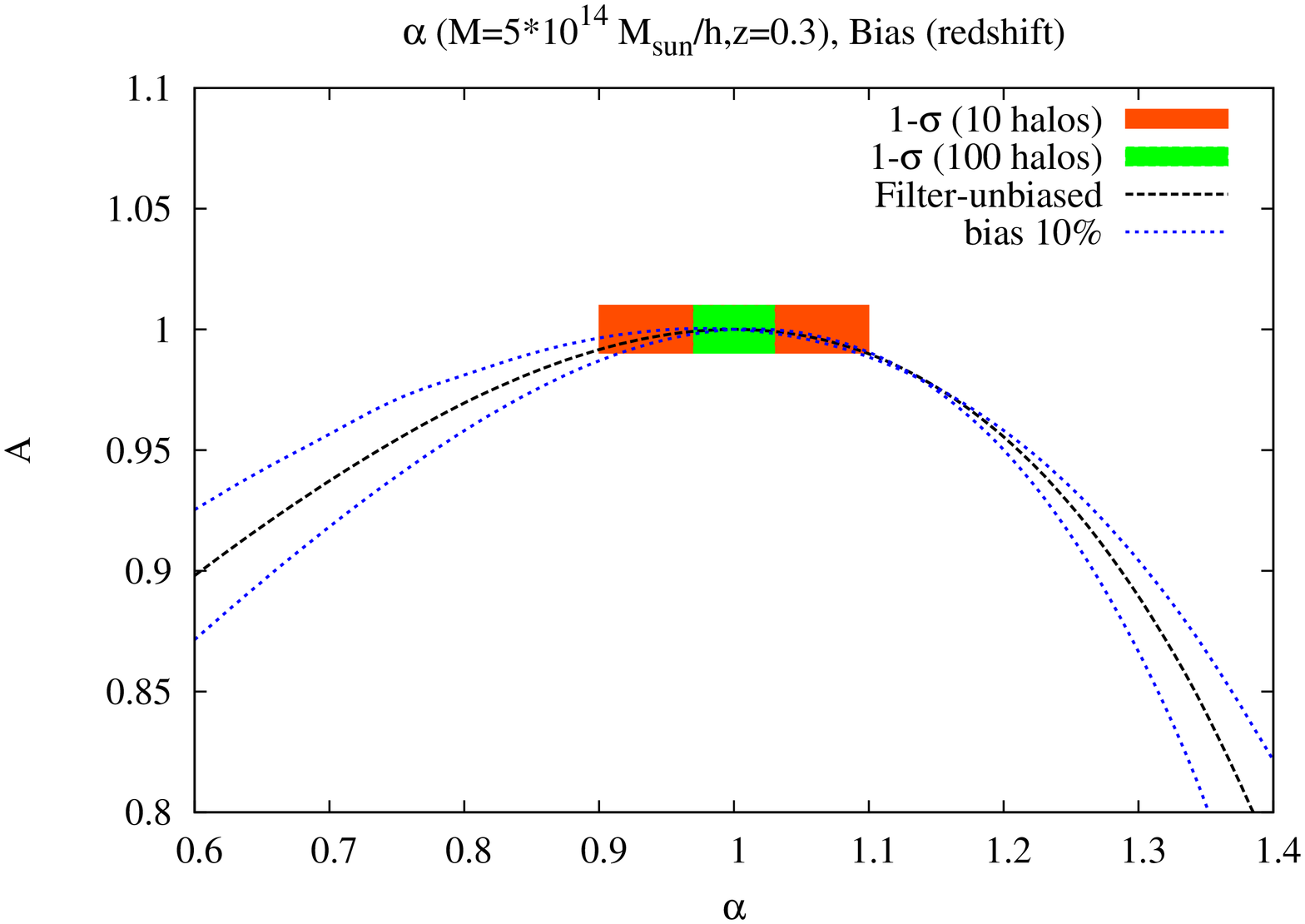}} &
      \resizebox{70mm}{!}{\includegraphics[width=7.0cm,angle=270]{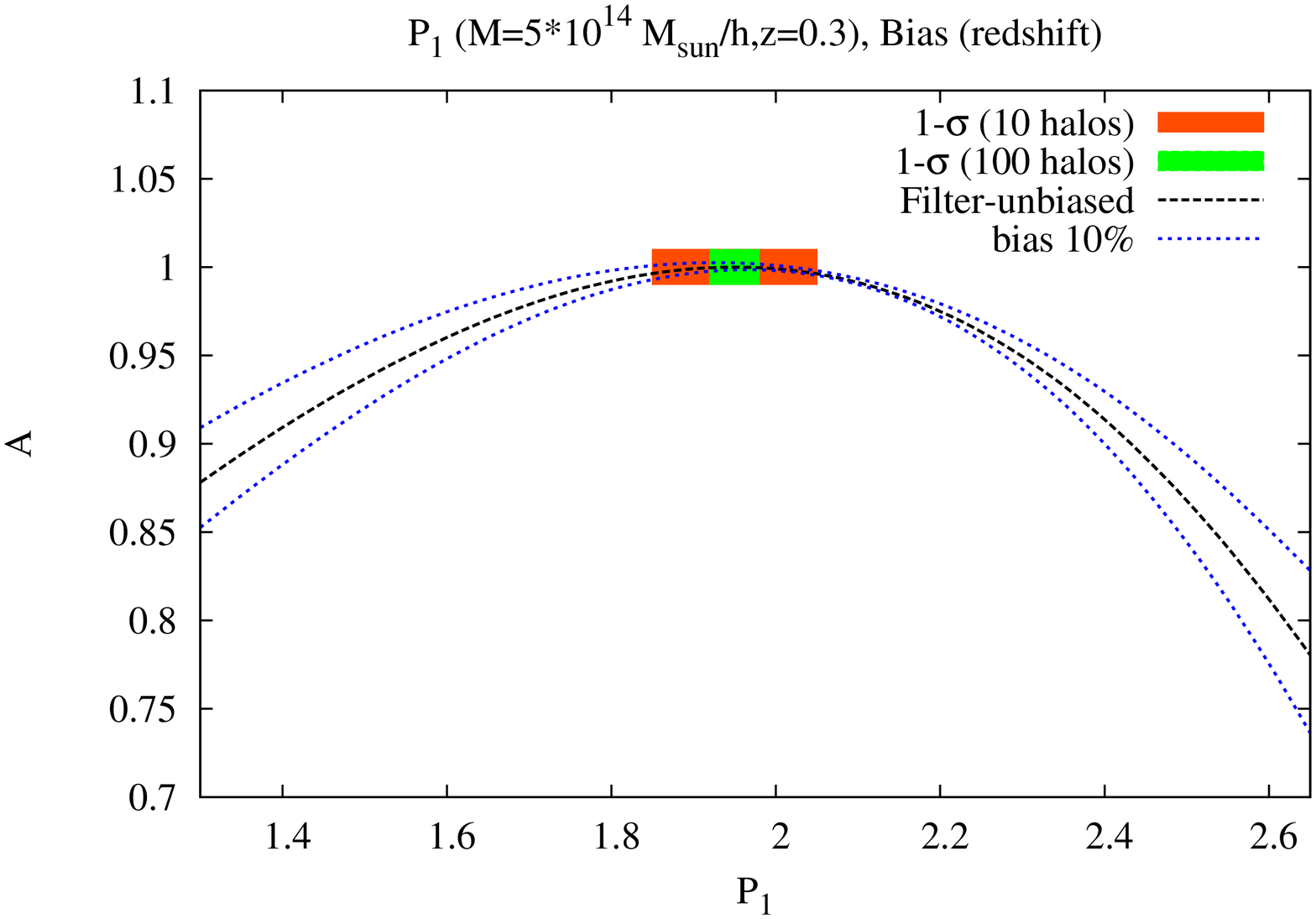}} \\
    \end{tabular}
    \caption{Estimated shear amplitude (Eq.~\ref{aest_fltr}) normalised to unity as a function of the inner slope ($\alpha$) (\textbf{left panels}) and as a function of $P1$ used in the filter definition (\textbf{right panels}). The \textit{black dashed curve} represents the case in which fiducial concentration,  mass and redshift used in the filter are correct. The \textit{blue curves} represent the effect of defining the filter with a wrong fiducial value for the concentration (first panel), the mass (second panel) and the redshift (third panel). The errors on the measurement of $\alpha$ were computed by Monte-Carlo simulation (see the text for details) and are rescaled for 10 and 100 haloes.}\label{fig:matched}
    \end{minipage}
  \end{center}
\end{figure*}

Defining the filter requires the specification of a model. The estimator (Eq.~\ref{aest_fltr}) we defined for the inner slope is unbiased only if the model is correct. We investigate here what happens if the filter is defined using a generalised NFW profile with wrong fiducial values of mass, redshift and concentration. We study in particular the case in which the fiducial redshift used in the filter differs from the real redshift by about 10~\%, the mass by about 30~\%, and the concentration by about 20~\%. We show the results in the first three panels of Figs.~\ref{fig:alpha_lin} and \ref{fig:matched} (blue lines) for the linear and the scale-adaptive filter, respectively.

As expected, the inner-slope estimate is biased. This reflects the degeneracy between the parameters, in particular between the scale radius $r_{-2}=r_{200}(M,z)/c_{-2}$ and the inner slope. The scale radius depends only slightly on the halo mass and redshift, while it is strongly affected by a variation in the concentration.

This bias has to be compared with the statistical errors associated with the measurement in order to assess whether uncertainties in the fiducial halo parameters are important or not. If a single halo is considered, a wrong assumption on the concentration (the most critical parameter) introduces a bias that is on the same order as the statistical error. However, if several haloes are stacked (we show in Fig.~\ref{fig:alpha_lin} results after stacking 10 and 100 haloes), the bias is a factor of 10 larger than the statistical uncertainty.

In Sect.~\ref{sec:degeneracy}, we discussed how it is possible to deal with degeneracies between inner slope and concentration, defining two new parameters ($P_1$, $P_2$), linear combinations of $c$ and $\alpha$, which are respectively the best and the worst constrained parameters given our model. The measurement of the new parameter $P_1$ is almost unaffected by the choice of the other parameter $P_2$ as we show in the right panel of Fig.~\ref{fig:alpha_lin} and \ref{fig:matched}, while the effect of a wrong assumption of halo mass and redshift produces a similar bias. We recall that these latter quantities can be measured by means of other observables, as discussed before.

Once the model had been re-parametrised in term of $P_1$ and $P_2$, we estimated the error on $P_1$ using a Monte-Carlo simulation in the same way we have done before for $\alpha$. The result is shown in Fig.~\ref{fig:Montecarlo_p1}.

\section{Potential problems}

We now want to point out the conditions under which the two methods described can be successfully applied.

First of all, the reduced shear must be measurable at relatively small angular scales (smaller than the scale radius of the halo) where the density profile is sensitive to a change of the inner slope.

\begin{figure}
\begin{center}
\hbox{
\psfig{figure=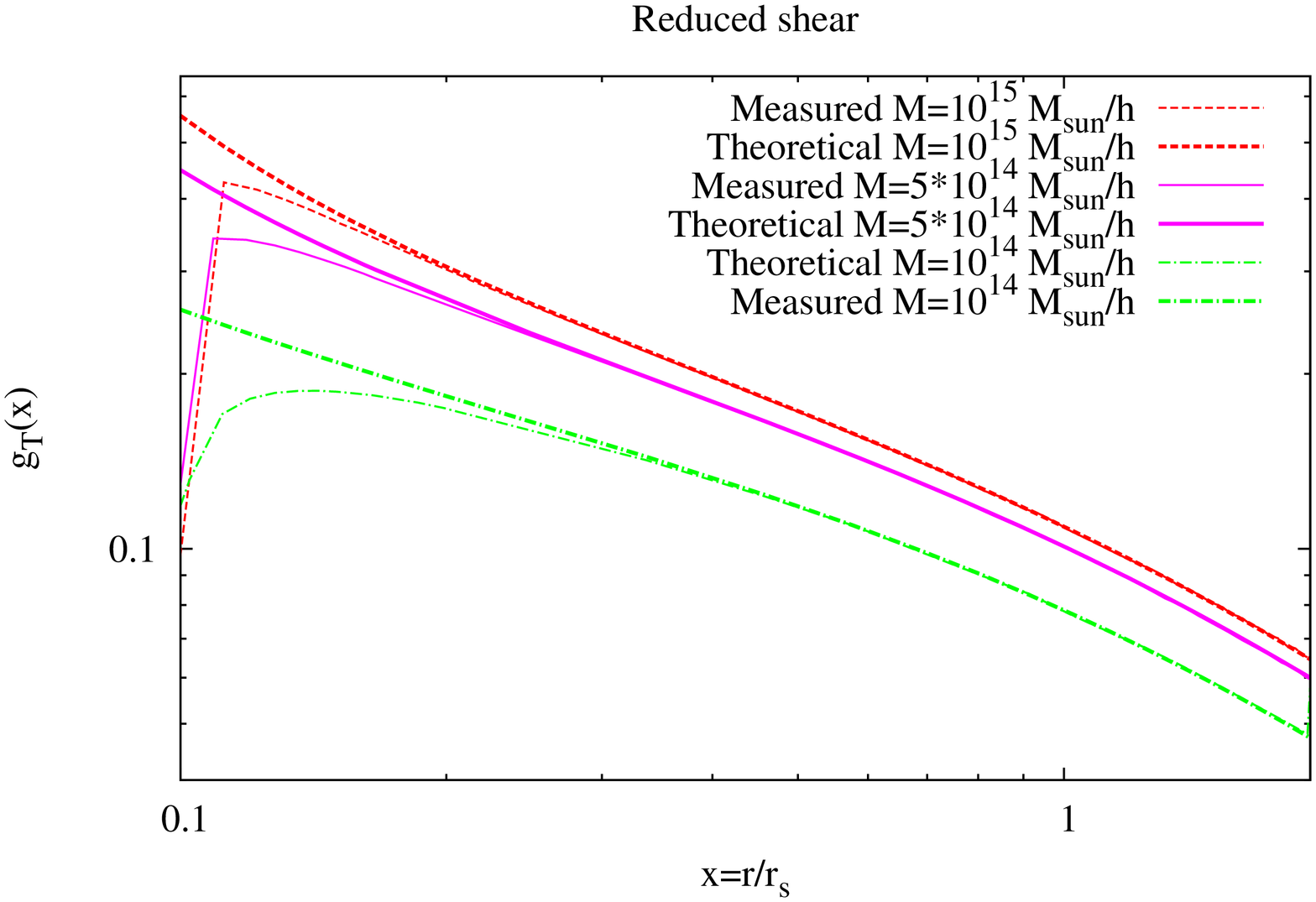,width=9cm,angle=270}
}
   \caption{Comparison between the theoretical reduced tangential shear (thick lines) and the shear estimated from galaxy ellipticities (thin lines) for three different masses.}\label{fig:UnbShear}
   \end{center}
\end{figure}

Towards the halo's centre, the image distortion becomes non-linear such that the galaxy ellipticities are no longer an unbiased estimator of the shear. We quantify the expected deviation by a simple test: We use the deflection-angle map of an NFW halo to lens a circular source (for which we assumed a Sersic profile with $n=1.5$ and $r=0.35 arcsec$) moving radially towards the halo centre. We measure the ellipticity of its image (using quadrupole moments) as a function of cluster-centric distance and compare it to the true reduced shear. Figure \ref{fig:UnbShear} shows the result for three different haloes ($M=10^{14}, 5\times 10^{14}, 10^{15} M_{\odot}/h$). The conclusion is that up to $r=0.2 r_{s}$ the measured ellipticity of galaxies is still an unbiased estimator of the (reduced) shear while at smaller scales the contribution from higher order terms start to be dominant. Therefore, $r\approx0.2r_s$ should be taken as the minimum radius where the measured ellipticity can still be considered to faithfully represent the reduced shear.

However measuring shear at these scales can be tricky even with a high background galaxies density due to the possible dilution of the shear signal caused by cluster galaxies. In order to avoid this problem, accurate colour-magnitude information should be available so that it is possible to well separate cluster members from non-members \citep{Broadhurst05}.

\begin{figure}
\begin{center}
\hbox{
\psfig{figure=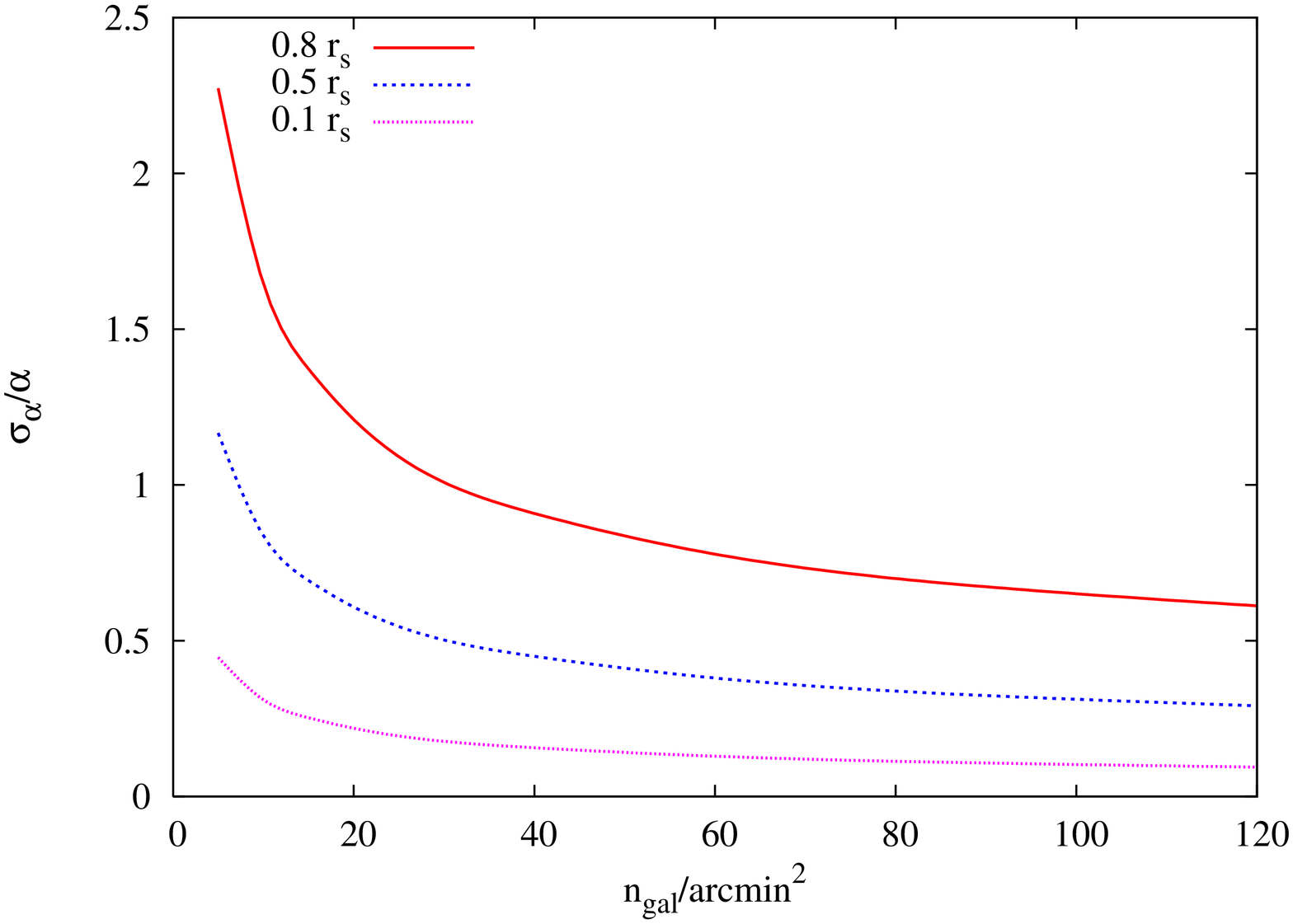,width=9cm,angle=270}
}
   \caption{Standard deviation of the inner slope as a function of the background-galaxy number density and of the minimal radius where the shear can be detected. The calculation has been done for a halo of $M=5\times 10^{14} M_{\odot}/h$ at redshift $z=0.3$ with $\alpha=1.0$ using the linear filter.}\label{fig:SN_ngal}
   \end{center}
\end{figure}

We showed in the previous section that the error associated to the measurement of the inner slope is high when computed for a single halo. Thus several haloes need to be stacked together. The number of haloes to be stacked depends strongly on the minimal radius where the shear can be detected, and on the number of background galaxies. In Fig.~\ref{fig:SN_ngal}, we plot the relative error on the measurement of $\alpha$ as a function of these two parameters for a halo of $M=5\times10^{14} M_{\odot}/h$ at redshift $z=0.3$. Assuming 30 galaxies per square arc minute, the number of haloes to be stacked to reach an accuracy of a few percent on the inner slope is between 10 and 100 going from $r_{min}=0.2 r_{s}$ to $r_{min}=0.8 r_{s}$. We emphasise that the stacking procedure can be affected by a wrong determination of the cluster centre that causes a circularisation of the average cluster profile in its central part \citep{Kath08}.  

\cite{Meneghetti07} showed how the determination of the inner slope can be biased if the triaxiality structure of the haloes are not taken properly into account. However if many haloes are stacked together a direct comparison with the projected DM average profile found using stacked simulated clusters can be consistently done.
   
Moreover the effect of the baryons in shaping the density profile at this scale is not negligible. We plan to attack this problem using numerical simulation in order to study the effect of stacking and the presence of the baryons on our results.

\section{Conclusion}

Starting from the question how the central density profiles of group or cluster-sized, DM haloes can best be constrained and compared to observations, we have developed two methods based on linear filtering of gravitational-shear data that aim at returning a single number, i.e.~an estimate of the inner slope $\alpha$ of density profile. One filter is constructed to directly return this number, the other searches for the maximum of the signal-to-noise ratio as a function of $\alpha$. Our results are as follows:

\begin{itemize}

\item When applied to a single halo of $5\times10^{14}\,M_{\odot}/h$ near $z=0.3$, the inner slope of the density profile can be estimated with a 1-$\sigma$ accuracy of $14\,\%$ with the linear filter and $19\,\%$ with the scale-adaptive filter, provided the halo concentration is known. Even though this situation is unrealistically idealised, it is promising because it is based on a single halo only.

\item Taking the considerable uncertainty in halo concentrations into account increases the 1-$\sigma$ error to between $25\ldots30\,\%$.

\item In reality, the halo concentration is at best roughly known. Based on real data, there is an almost perfect degeneracy between $\alpha$ and the halo-concentration parameter $c$: if $c$ is assumed to be too large, $\alpha$ will be underestimated and vice versa. Based on lensing data only, this degeneracy cannot be lifted.

\item To address this problem, we search for that combination of the parameters $\alpha$ and $c$ that can best be constrained by observations. We set up the Fisher matrix, rotate the two-dimensional parameter space to diagonalise it and identify its smaller eigenvalue as that best-constrained parameter, called $P_1$. We find $P_1=0.95\alpha+ 0.30 c$ for the linear filter and $P_1=0.97 \alpha+ 0.22 c$ for the scale-adaptive filter. 

\item This parameter $P_1$ is now constrained with a 1-$\sigma$ relative accuracy of $\sim14\,\%$ both with the linear and the scale-adaptive filters and the measurement is almost insensitive to the value of the other parameter $P_2$.

\end{itemize}

While these results seem highly promising, in particular when applications to cluster samples rather than individual clusters are envisaged, we consider our study as a first step. While we have taken into account that image ellipticities measure the reduced gravitational shear rather than the shear itself, measuring the reduced shear near the centres of galaxy groups or clusters is severely hampered by the cluster galaxies themselves. It thus appears necessary to stack the signal from several or many clusters to arrive at a reliable estimate for $\alpha$. Then, clusters with different masses, redshifts and concentration parameters will inevitably be combined, with the tendency to blur the signal. However, the results derived and presented above indicate that the principle of our approach is promising, which consists in combining all available information into a single number, which is thus well constrained. Further studies are required to address the issue of stacking data in this context.

\section*{Acknowledgements}

This work was supported by the EU-RTN ``DUEL'', the Heidelberg Graduate School of Fundamental Physics, the IMPRS for Astronomy \& Cosmic Physics at the University of Heidelberg and the Transregio-Sonderforschungsbereich TR~33 of the Deutsche Forschungsgemeinschaft. We thank Massimo Meneghetti for useful comments and criticisms and Peter Melchior for fruitful discussion about shear measurement in galaxy clusters.

\bibliographystyle{mn2e}

\end{document}